\documentclass[12pt]{article}
\pdfoutput=1
\usepackage{jheppub}
\usepackage[utf8]{inputenc}
\usepackage{verbatim}
\usepackage{amsmath}
\usepackage{amssymb}
\usepackage{amsthm}
\usepackage{amsfonts}
\usepackage{hyperref}
\usepackage{xcolor}
\newcommand{\be}{\begin{equation}}
\newcommand{\ee}{\end{equation}}
\newcommand{\ran}{\rangle}
\newcommand{\lan}{\langle}

\newcommand{\ol}{\overline}
\newcommand{\wt}{\widetilde}
\newcommand{\bi}{\begin{itemize}}
\newcommand{\ei}{\end{itemize}}
\newcommand{\Tr}{\mathrm{Tr}}
\newcommand{\bfig}{\begin{figure}\begin{center}}
\newcommand{\efig}{\end{center}\end{figure}}
\newcommand{\C}{\mathcal{C}}
\newcommand{\R}{\mathcal{R}}
\newcommand{\T}{\mathcal{T}}
\newcommand{\RT}{\mathcal{RT}}
\newcommand{\CRT}{\mathcal{CRT}}

\usepackage{graphicx}
\begin{document}
\title{Gauging spacetime inversions in quantum gravity}
\author[a]{Daniel Harlow}
\author[b]{and Tokiro Numasawa}
\affiliation[a]{Center for Theoretical Physics\\ Massachusetts Institute of Technology, Cambridge, MA 02139, USA}
\affiliation[b]{Institute for Solid State Physics,  University of Tokyo, Kashiwa, Chiba 277-8581, Japan}
\emailAdd{harlow@mit.edu, numasawa@g.ecc.u-tokyo.ac.jp}
\abstract{Spacetime inversion symmetries such as parity and time reversal play a central role in physics, but they are usually treated as global symmetries.  In quantum gravity there are no global symmetries, so any spacetime inversion symmetries must be gauge symmetries.  In particular this includes $\mathcal{CRT}$ symmetry (in even dimensions usually combined with a rotation to become $\mathcal{CPT}$), which in quantum field theory is always a symmetry and seems likely to be a symmetry of quantum gravity as well.  In this article we discuss what it means to gauge a spacetime inversion symmetry, and we explain some of the more unusual consequences of doing this.  In particular we argue that the gauging of $\mathcal{CRT}$ is automatically implemented by the sum over topologies in the Euclidean gravity path integral, that in a closed universe the Hilbert space of quantum gravity must be a real vector space, and that in Lorentzian signature manifolds which are not time-orientable must be included as valid configurations of the theory. In particular we give an example of an asymptotically-AdS time-unorientable geometry which must be included to reproduce computable results in the dual CFT.}
\maketitle

\section{Introduction}
For centuries it was believed that the laws of physics are invariant under both spatial reflection and time-reversal.  This is now understood to be false due to the chiral nature of the electroweak interactions, but there is a particular combination of spatial and temporal reflection, called $\C\R\T$, which is provably a symmetry of any relativistic quantum field theory \cite{Luders:1954zz,pauli1988exclusion,Streater:1989vi,Weinberg:1995mt}.  The reason for $\C\R\T$ symmetry in relativistic quantum field theories is easy to explain: $\C\R\T$ is what we get by performing a rotation by $\pi$ in the plane of one spatial direction and time in Euclidean signature, which inverts both directions, and then analytically continuing to Lorentzian signature.\footnote{The name $\C\R\T$ indicates that this symmetry takes the complex $\C$onjugate of fields, $\R$eflects one spatial direction, and implements a $\T$ime reversal.  This can be misleading however, as none of $\C$, $\R$, or $\T$ need be symmetries separately, and even when they are there are ambiguities in their definitions since they can mix with internal symmetries in nontrivial ways.  By contrast $\C\R\T$ is uniquely defined as a rotation by $\pi$ in Euclidean signature.  In even spacetime dimensions one traditionally combines $\C\R\T$ with a spatial rotation such that all spatial coordinates are reflected rather than just one, with the combined symmetry being called $\C\mathcal{P}\T$. This combination does not work in odd dimensions however, and in any case the proof of the theorem only reflects one spatial direction so we might as well respect this.  Despite efforts to the contrary \cite{Florida}, $\CRT$ still applies in the state of Florida.}  Since Lorentz symmetry becomes rotational invariance in Euclidean signature, this operation is a symmetry in any relativistic quantum field theory that can be analytically continued to Euclidean signature (which is the case provided that the Hamiltonian is bounded from below).

In this paper our primary interest is the role of spacetime inversion symmetries in quantum gravity.  By now it is widely expected that in quantum gravity there are no global symmetries \cite{Zeldovich:1976vq,Zeldovich:1977be,Giddings:1988cx,Abbott:1989jw,Coleman:1989zu,Kallosh:1995hi,Banks:2010zn,Harlow:2018jwu,Harlow:2018tng,Harlow:2020bee,Chen:2020ojn,Hsin:2020mfa,Bah:2022uyz}.  In particular it was argued in \cite{Harlow:2018tng} that this conclusion applies to spacetime inversions.  One possibility is that spacetime inversions are simply not symmetries in quantum gravity, as indeed seems to be the case for $\R$ and $\T$ in our world.  On the other hand we know that $\CRT$ is a symmetry of all relativistic field theories, and it is quite plausible that it must also be a symmetry of quantum gravity.  This is certainly true within the AdS/CFT correspondence, since there quantum gravity in asymptotically-AdS space is formulated as a relativistic quantum field theory on the boundary and thus has $\CRT$ symmetry. More generally $\CRT$ is a perturbative symmetry of all known string vacua, with the proof being a rotation by $\pi$ in the target space of the Euclidean string \cite{Polchinski:1998rr}.  Assuming that $\CRT$ is indeed a symmetry of quantum gravity, we are led to the conclusion that it must be a gauge symmetry since these are not forbidden.  Moreover in vacua where versions of $\R$ and $\C\T$ separately are symmetries (given $\CRT$ symmetry it is both or none), they must be gauged as well.

The goal of this paper is therefore to understand what it means for a spacetime inversion to be a gauge symmetry.  It is not immediately clear that this makes sense, especially when the symmetry involves $\T$.  More specifically we are confronted with the following issues:
\bi
\item If we have a gauge symmetry that reverses time, then there are valid spacetime configurations with the property that moving around a spatial circle reverses the direction of time.  In other words there are valid spacetimes that are not time-orientable. How are we to make sense of quantum field theory in such spacetimes?  Why do they not lead to unacceptable causal pathologies?
\item In a closed universe all physical states should be gauge invariant.  In quantum mechanics any symmetry that reverses time is represented by an antiunitary operator $\Theta$ on Hilbert space,\footnote{The usual argument for this statement is that if $\Theta$ were unitary then  to have $\Theta^\dagger e^{-iHt}\Theta=e^{iHt}$ we would need to have $\Theta^\dagger H \Theta=-H$, which implies that the spectrum of the Hamilonian is symmetric about zero.  Since this is usually not the case, we instead take $\Theta$ to commute with the Hamiltonian but be antiunitary - the desired action on $e^{-iHt}$ then arises because $\Theta i=-i\Theta$. For dynamical gravity in a closed universe this argument is more subtle to make since the total Hamiltonian is zero, but if we assume that the matter theory in a fixed background does not have a symmetric spectrum then the conclusion carries over.} and the states which are invariant under an antiunitary operator form a real vector space.  For example if $\Theta|\psi\ran=|\psi\ran$ then $\Theta i|\psi\ran=-i|\psi\ran$.  Since $\CRT$ is a gauge symmetry that reverses time, we learn that the Hilbert space of quantum gravity in a closed universe is a real vector space.  How is this compatible with our conventional understanding of quantum mechanics, where the Hilbert space is always complex?  
\item The Lorentz group has four connected components, which are those containing the identity, $\R$, $\T$, and $\RT$, while the Euclidean rotation group has only two (we review these statements in section \ref{topsec}).  This leads to a puzzling situation.  From the point of view of the gravitational path integral there is no ban on global symmetries.  Given a low-energy gravitational Lagrangian that has $\R$ (and thus also $\C\T$) symmetry, we therefore have five choices about which spacetime inversions to gauge: we could gauge none of $\R$, $\C\T$, or $\CRT$, we could gauge all of $\R$, $\C\T$, and $\CRT$, we could gauge $\R$ but not $\C\T$, we could gauge $\C\T$ but not $\R$, or we could gauge $\CRT$ but not $\R$ or $\C\T$ separately.   In Euclidean signature on the other hand it seems that the only choice is whether to gauge the non-identity component of the rotation group or not.  What happened to the other choices? Whichever choices are allowed, how should we interpret them in a true non-perturbative theory of quantum gravity that doesn't allow global symmetries?
\ei
On the other hand in Euclidean signature there is a long history of gauging spacetime inversions, going back to the gauging of worldsheet parity in unoriented string theory \cite{Polchinski:1998rr}.  More recently the gauging of $\T$ symmetry has turned out to be an important part of the connection between Euclidean JT gravity and random matrices \cite{Saad:2019lba,Stanford:2019vob} (we discuss these examples in more detail in sections \ref{backsec} and \ref{conc} respectively).  We therefore need to address the above issues in a convincing way, and we hope to do so with this paper.  Roughly speaking our answers are as follows:
\bi
\item Quantum field theory on a spacetime that is not time-orientable does make sense, provided that we are careful to follow the rules for working on a nontrivial background for a global symmetry.  In field theory the causal behavior on such a background is indeed strange, but in gravity we can hope that such behavior is hidden behind horizons due to something we call \textit{chronological censorship}.  More concretely, we conjecture that all self-intersecting timelike curves (SITCs) have their self-intersections hidden behind horizons.  At least in AdS/CFT this is apparently the case, since boundary causality must be exactly preserved.  One could try to use boundary causality as an argument that geometries with SITCs shouldn't be included in the first place, but we are able to give an explicit example of an asymptotically-AdS spacetime that is not time-orientable, and which has SITCs, but that nonetheless needs to be included to match a calculation in the boundary CFT.  This confirms that $\CRT$ is indeed a bulk gauge symmetry in AdS/CFT.
\item There is no insurmountable problem with the Hilbert space of a closed universe being real, and we are able to recover all the usual predictions of quantum mechanics in such a universe in the limit that it is big.  Essentially this works in the same way as electrodynamics in a closed universe, where locally we can see objects that are charged even though the total charge must be zero.  This reality of the Hilbert space is however a strong constraint on any fundamental theory of quantum gravity in a closed universe, and may thus be an aid in finding one.
\item In Euclidean quantum gravity the gauging of $\CRT$ is automatically imposed by the standard sum over topologies in the path integral.\footnote{In an interesting recent paper, Witten explains how a $\CRT$-like symmetry he calls ``$\T$'' can be treated as global if one ``decorates'' the Euclidean path integral by assigning an orientation to each closed boundary component that need not coincide with the induced orientation from the bulk \cite{Witten:2025ayw}.  This allows one to define a Euclidean path integral where this $\T$ is a global symmetry, albeit with the unusual feature that the value of the path integral is independent of this choice of boundary orientation.  The orientations in this approach have meaning only when one assigns a Hilbert space interpretation to the path integral, in which case they tell us which boundary components are bras and kets.  Our expectation is still that in holographic theories $\CRT$ must be gauged, in particular in holography boundary data should come from a bulk field but there seems to be no bulk field for the boundary orientations in Witten's approach.  We will also see in section \ref{adssec} below that nontrivial $\CRT$-bundles must be allowed in the bulk to match CFT calculations.  }  This therefore eliminates most of the apparent choice in the gauge group, with the only remaining choices being for $\R$, $\C\T$, and $\CRT$ to all be gauged or for $\CRT$ to be gauged but $\R$ and $\C\T$ separately to be global.  In Euclidean signature the former amounts to including unoriented Euclidean geometries in the path integral, while the latter excludes them.  Other choices of gauging in Lorentzian signature amount to nonlocal restrictions on the sum over Euclidean topologies.  When $\R$ and $\C\T$ are global one can try to interpret them as symmetries of an ensemble of gravitational theories that break them in any particular instantiation, with the relationship between JT gravity and random matrix theory giving an explicit example, but in higher dimensions it is not clear if such an ensemble interpretation makes sense and most likely $\R$ and $\C\T$ should just be broken.  
\ei

Our plan for the rest of this paper is the following: in section \ref{intsec} we review what it means to gauge a discrete internal symmetry.  In section \ref{topsec} we review the topology of the Lorentz group and the $\CRT$ theorem.  In section \ref{backsec} we explain how to turn on background gauge fields for spacetime inversions in quantum field theory, and argue that this does not lead to any inconsistencies.  In section \ref{sumsec} we explain how the gauging of spacetime inversions is implemented in bosonic theories via the sum over topologies in the path integral. In section \ref{adssec} we present an example of a nontrivial $\CRT$ bundle that appears in asymptotically-AdS space, and we demonstrate that this spacetime is needed to reproduce CFT results.  In section \ref{censorsec} we give a general argument for chronological censorship.  In section \ref{quantsec} we further discuss the idea that the Hilbert space of quantum gravity in a closed universe is a real vector space, in particular explaining why this is consistent with the observed consequences of quantum mechanics and also how it addresses an old puzzle in quantum cosmology.  In section \ref{fermsec} we consider how to generalize the discussion of section \ref{sumsec} to include fermions, emphasizing in particular that in gravity fermion parity is always a gauge symmetry.  Finally in section \ref{conc} we discuss how the results of this paper relate to some other recent ideas in the literature.

\section{Gauging discrete internal symmetries}\label{intsec}
As a warmup it is convenient to recall what it means to gauge a discrete internal symmetry.  Our presentation begins in a somewhat formal way, so some readers may want to jump to the examples in section \ref{intexsec}.  

\subsection{General discussion}
There are various perspectives on how gauge and global symmetries in quantum field theory should be defined. The approach we will adopt here is that of \cite{Harlow:2018tng}, which defines an internal global symmetry with symmetry group $G$ in quantum field theory as a family of unitary codimension-one surface operators $U(g,\Sigma)$ acting on the Hilbert space of theory quantized on  each spatial Riemannian manifold $\Sigma$ such that\footnote{There is also a technical continuity requirement for the symmetry action, see \cite{Harlow:2018tng} for the details.  In the language of the generalized global symmetries of \cite{Gaiotto:2014kfa} what we are defining here is a zero-form global symmetry; we will not need to discuss higher-form symmetries in this paper.}
\begin{enumerate}
\item The group algebra is respected:
\be
U(g_1,\Sigma)U(g_2,\Sigma)=U(g_1g_2,\Sigma).
\ee
\item The local algebra $\mathcal{A}[R]$ of operators associated to any spatial region $R\subset \Sigma$ is preserved by the symmetry:
\be
U(g,\Sigma)^\dagger \mathcal{A}[R]U(g,\Sigma)=\mathcal{A}[R] \qquad \forall g\in G
\ee
\item This action of $G$ by conjugation is faithful on the set of local operators, and moreover the representation $D_{ij}(g)$ defined by 
\be
U^\dagger(g,\Sigma)O_i(x)U(g,\Sigma)=\sum_jD_{ij}(g)O_j(x)
\ee
where $O_i$ is a basis of local operators, is independent of the choice of $\Sigma$.
\item The energy-momentum tensor is invariant:
\be
U^\dagger(g,\Sigma)T_{\mu\nu}(x)U(g,\Sigma)=T_{\mu\nu}(x).
\ee
\end{enumerate}
The second requirement ensures that the symmetry action is compatible with the local structure of the theory, sending local operators to local operators at the same point, line operators to line operators on the same line, and so on.  The third requirement ensures we have correctly identified the (zero-form) symmetry group.  The fourth requirement is equivalent to saying that the codimension-one surface on which $U(g,\Sigma)$ is supported can be continuously deformed in correlation functions without changing them as long as it doesn't intersect another operator.  See \cite{Harlow:2018tng} for more motivation and discussion of this definition.  In what follows we will usually shorten $U(g,\Sigma)$ to $U(g)$, but it should be remembered that in the path integral $U(g)$ is not fully specified until we say what codimension-one surface it lives on.

An important operation in the context of global symmetries is turning on a background gauge field.  
The mathematical definition of a background gauge field for a global symmetry with symmetry group $G$ on a spacetime manifold $M$ is a connection on a principal $G$-bundle with $M$ as its base space (see e.g. \cite{Nakahara:2003nw}). When $G$ is a discrete group this definition simplifies: a background gauge field on $M$ is a choice of homomorphism 
\be
w:\pi_1(M,x_0)\to G,
\ee
where $\pi_1(M,x_0)$ is the fundamental group of $M$ with basepoint $x_0$ (here we assume that $M$ is connected so the choice of $x_0$ is arbitrary), and we view two such homomorphisms $w_1,w_2$ as equivalent if $w_1=g_0w_2g_0^{-1}$ for some $g_0\in G$.\footnote{The idea behind this simplification is that when $G$ is discrete, any principal $G$-bundle $P$ over $M$ is a covering space.  These are classified by how the elements of the fiber $G$ are permuted as we move around each closed loop in $M$ starting at a point in $P$, and for a principal $G$-bundle this action must be a right-multiplication on the fiber by some element of $G$ (see e.g. section 1.3 of \cite{hatcher}).}  The meaning of $w$ is that if we move a local operator $O$ transforming in a representation $\alpha$ of $G$ around a closed loop $C$, then $O$ transforms as
\be
O\to D_\alpha(w([C]))O
\ee
where $[C]$ is the homotopy class of $C$ and $D_\alpha(w([C]))$ is called the holonomy.  The identification of conjugate homomorphisms is needed because we can do a gauge transformation by $g_0\in G$ at the base point $x_0$, and this causes all the holonomies be conjugated by $g_0$ (this is why we usually take the trace of the holonomy to define the gauge-invariant Wilson loop).  

\bfig
\includegraphics[height=3cm]{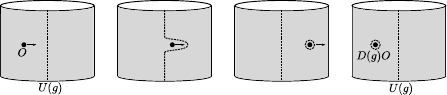}
\caption{Moving a charged operator $O$ around a closed path in the presence of a background gauge field for a discrete global symmetry.  As $O$ crosses the symmetry insertion $U(g)$ it picks up a transformation matrix $D(g)$.}\label{cyclefig}
\efig
In practice one implements a discrete background gauge field as follows: for each loop $C$ in $M$ we can choose a Poincare dual $(d-1)$-cycle $\wt{C}$, and then we insert $U(w([C]))$ on this dual cycle.  The holonomy action on local operators arises as in figure \ref{cyclefig}.  We emphasize that in this construction the correlators of the theory \textit{know} the location of the symmetry insertion $U(g)$: its location is the surface where any charged operator $O$ discontinuously picks up a transformation $D(g)$ as we move it across the surface.  This means that there is more information in the locations of the $U(g)$ than there is in the homomorphism $w$. This information however is lost once we make the gauge field dynamical, as we will now see.

To gauge our global symmetry $G$, meaning to make the discrete gauge field dynamical, we sum over the homomorphisms $w$ in the path integral and divide by the size of $G$.  Since we have realized these homomorphisms by placing $U(g)$ insertions on $(d-1)$-cycles, we are really summing over networks of $U(g)$ insertions.\footnote{In general we cannot do this sum consistently unless there is a rule for dealing with intersections of $U(g)$ insertions.  In particular for each crossing there could be more than one way to ``resolve'' the crossing, and if these resolutions do not agree then the symmetry has an 't Hooft anomaly and cannot be gauged \cite{Lin:2019kpn}. This is the discrete analogue of the familiar idea that perturbative anomalies such as the chiral anomaly come from contact terms in the current correlation functions.  In this paper we are interested in discrete symmetries that can be gauged, so we will assume they have no such anomalies.}  It is useful to consider the Hilbert space interpretation of this gauging for a field theory on a spatial manifold $\Sigma$.  $U(g)$ is codimension-one, so it can either be wrapping all of $\Sigma$ or it can be wrapping some $(d-2)$-cycle of $\Sigma$ and be extended in the time direction.  In the former case the sum over $g$ serves to enforce a projection onto $G$-invariant states, as we would expect in a gauge theory.  When $U(g)$ is extended in the time direction we should think of it as a defect that introduces a new sector of the Hilbert space, conventionally called a twisted sector.  There is a twisted sector for each homomorphism $r:\pi_1(\Sigma)\to G$ modulo conjugation by an element of $G$.  Since we can also have $U(g)$ wrapping $\Sigma$ in each twisted sector, we need to project onto gauge-invariant states in each twisted sector as well.   This restriction to gauge-invariant states (and thus also gauge-invariant observables) has a very important consequence for what follows: it means that once the symmetry is gauged, the correlation functions are no longer sensitive to deformations of the surfaces on which the $U(g)$ insertions are located.  The charged operators that we could have used to detect this are no longer present, and so we can place the $U(g)$ insertions in the sum over $w$ wherever we like (as long as we include one dual cycle for each element of $\pi_1(M,x_0)$).

\subsection{Examples of discrete gauging}\label{intexsec}
Let's illustrate this general discussion using a pair of simple examples in $1+1$ dimensions.  The first will be the $\phi'=-\phi$ symmetry in the theory of a free massless scalar with Lagrangian
\be
\mathcal{L}=-\frac{1}{2}\partial_\mu \phi \partial^\mu \phi.
\ee
To avoid trouble with the zero mode we will take the target space of this scalar to be a circle of radius $R_\phi$:
\be
\phi\sim \phi+2\pi R_\phi.
\ee
We can illustrate the gauging of $\phi'=-\phi$ symmetry in this theory using its thermal partition function on a spatial circle of circumference $L$.  If $\phi'=-\phi$ is a \textit{global} symmetry, then the thermodynamic partition function is given by  
\be\label{Z1bos}
Z_1=\Tr\left(e^{-\beta H+i\theta J}\right).
\ee
Here we have included a spatial rotation by $\theta$ in the trace for future reference ($J$ is the angular momentum on the spatial circle). The spatial periodicity implies the boundary condition 
\be
\phi(t,x+L)=\phi(t,x)+2\pi R_\phi m,
\ee
where $m$ is the winding number of the scalar in target space as we move around the spatial circle.  In computing $Z_1$ we need to sum over $m$ (see appendix \ref{freeapp} for a review of the calculation of $Z_1$).

\bfig
\includegraphics[height=3cm]{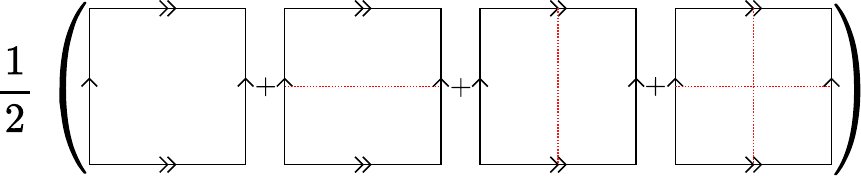}
\caption{Gauging an internal $\mathbb{Z}_2$ symmetry in $1+1$ dimensions.  We need to sum over trivial holonomy, a $\mathbb{Z}_2$ holonomy around the temporal direction, a $\mathbb{Z}_2$ holonomy around the spatial direction, and $\mathbb{Z}_2$ holonomies around around both directions.  The locations of the $U$ insertions implementing these holonomies are shown as red dashed lines, and for simplicity we have drawn a square torus with $\theta=0$.  The first two terms give a projection onto singlets in the untwisted sector, while the last two terms give a projection onto singlets in the twisted sector.}\label{gaugingfig}
\efig
If we instead treat $\phi'=-\phi$ as a gauge symmetry, then we need to introduce three new contributions to the partition function:
\begin{align}\nonumber
Z_2&=\Tr\left(U e^{-\beta H+i\theta J}\right)\\\nonumber
Z_3&=\Tr_U\left(e^{-\beta H+i\theta J}\right)\\
Z_4&=\Tr_U\left(U e^{-\beta H+i\theta J}\right).\label{bossect}
\end{align}
Here $U$ is the symmetry operator representing the nontrivial element of $\mathbb{Z}_2$, which acts on $\phi$ as
\be
U^\dagger\phi(t,x)U=-\phi(t,x).
\ee
$Z_2$ corresponds to a background gauge field with the symmetry operator wrapping a spatial circle, $Z_3$ corresponds to a background gauge field with the symmetry operator wrapping a temporal circle, and $Z_4$ corresponds to a background gauge field with symmetry operators wrapping both the spatial and temporal circles.  The symbol $\Tr_U$ indicates the trace in a twisted sector where $\phi$ obeys the antiperiodic boundary condition $\phi(t,x+L)=-\phi(t,x)+2\pi R_\phi m$.\footnote{The zero mode sector in the twisted sector is simpler than that of the untwisted sector, since there are no winding or momentum modes.  The meaning of $m$ in the twisted sector is simply that there are two ground states, one centered at $\phi=0$ and the other at $\phi=\pi R_\phi$, and we can build a Fock space of excitations on top of either one.} The full partition function is given by
\be
Z_{gauged}=\frac{1}{2}\left(Z_1+Z_2+Z_3+Z_4\right),
\ee
as shown in figure \ref{gaugingfig}.  The first two terms implement a projection onto gauge-invariant states in the untwisted sector, while the second two terms implement a projection onto gauge-invariant states in the twisted sector.  In string theory this gauged theory is called an orbifold.  Explicit formulas for $Z_1,\ldots, Z_4$ are given in appendix \ref{freeapp}.  

As a second example we can consider a free Majorana fermion in $1+1$ dimensions, which has Lagrangian
\be\label{IsingL}
\mathcal{L}=\frac{i}{2}\left[\psi_1 \left(\dot{\psi}_1-\psi'_1\right)+\psi_2\left(\dot{\psi}_2+\psi_2'\right)\right].
\ee
Here $\psi_1$ is left-moving and $\psi_2$ is right-moving.  The fermion algebra is
\begin{align}\nonumber
\{\psi_1(x),\psi_1(y)\}&=\{\psi_2(x),\psi_2(y)\}=\delta(x-y)\\
\{\psi_1(x),\psi_2(y)\}&=0.
\end{align}
We are interested in the fermion parity symmetry $(-1)^F$, which acts as $-1$ on both $\psi_1$ and $\psi_2$.  When this symmetry is a global symmetry, we need to impose periodic (also called Ramond) boundary conditions on $\psi_1$ and $\psi_2$.  The thermal trace is given by
\be
Z_1=\Tr_R\left(e^{-\beta H+i\theta J}\right),
\ee
which in the path integral picture sums over Euclidean fermion configurations that are periodic in the spatial direction but antiperiodic in the temporal direction due to the peculiarity of the fermion path integral:
\begin{align}\nonumber
\psi_i(t_E ,x+L)&=\psi_i(t_E ,x)\\
\psi_i(t_E +\beta,x)&=-\psi_i(t_E ,x).
\end{align}
When fermion parity is a global symmetry $Z_1$ is the full partition function.  We emphasize that since we are using different boundary conditions in the temporal and spatial directions,  $Z_1$ is not invariant under the $SL(2,\mathbb{Z})$ modular group of the torus.

To gauge fermion parity, we again need to sum over three new contributions to the partition function:
\begin{align}\nonumber
Z_2&=\Tr_R\left((-1)^F e^{-\beta H+i\theta J}\right)\\\nonumber
Z_3&=\Tr_{NS}\left(e^{-\beta H+i\theta J}\right)\\
Z_4&=\Tr_{NS}\left((-1)^F e^{-\beta H+i\theta J}\right).\label{fermsect}
\end{align}
Here $\Tr_{NS}$ is the trace over a twisted sector where the fermions obey antiperiodic boundary conditions in the spatial direction, which for historical reasons is called the Neveu-Schwarz sector.  The full partition function of the gauged theory is
\be
Z_{gauged}=\frac{1}{2}\left(Z_1+Z_2+Z_3+Z_4\right).
\ee
This partition function \textit{is} modular invariant, since we are now summing over a modular-invariant set of boundary conditions.  As before we can interpret the first two terms as the sum over gauge-invariant states in the untwisted (Ramond) sector and the last two terms as the sum over gauge-invariant states in the twisted (Neveu-Schwarz) sector.  We again give explicit formulas for $Z_1,\ldots,Z_4$ in appendix \ref{freeapp}.  

It is the gauged version of this Majorana fermion that describes the long-distance behavior of the critical Ising model. In particular the ``energy operator'' $\epsilon$ is given by $i\psi_1\psi_2$, while the Ising spin operator $\sigma$ is constructed by removing a small disk from the path integral and then imposing a boundary condition where $\psi_1$ and $\psi_2$ pick up a sign as we go around the edge of the disk (in modern parlance $\sigma$ is the Gukov-Witten operator associated to gauging fermion parity).  $\psi_1$ and $\psi_2$ themselves are not gauge-invariant, but we can construct gauge-invariant nonlocal operators by connecting any pair of these fermion operators at separate points by a Wilson line for fermion parity.  In Ising language this Wilson line is what is usually called the ``disorder line'', for example in the transverse field presentation of the model with Ising spin $\sigma_z$ the disorder line is a product of $\sigma_x$ operators along the line.\footnote{Somewhat counterintuitively the fermion parity Wilson line can also end on a scalar operator, which in Ising language is usually called the disorder operator.  In the language of the state-operator correspondence, the spin operator is the ground state in the Ramond sector which has even fermion parity while the disorder operator is the Ramond ground state which is odd.  The former is a genuine local operator on the gauge-invariant Hilbert space, while the latter can only live at the end of the fermion parity Wilson line.  The spin-statistics theorem is not violated since in the gauged theory all local operators are bosonic and have integer spin, while in the ungauged theory only the NS sector states correspond to local operators (the Ramond sector states correspond to nonlocal operators due to a branch cut in the conformal transformation from the cylinder to the plane). We thank Shu-heng Shao and Nati Seiberg for a useful discussion of this issue.} The fermion parity Wilson line is also the line that implements the usual Ising $\mathbb{Z}_2$ global symmetry $\sigma\to -\sigma$.  Fermion parity is also a gauge symmetry on the superstring worldsheet, where it gives rise to what is usually called the GSO projection.  These points were also recently discussed in \cite{Seiberg:2023cdc}.

\section{Topology of the Lorentz group and the $\CRT$ theorem}\label{topsec}
In this section we review some standard facts about $\CRT$ symmetry and the topology of the Lorentz group, along the way establishing some notation that will be useful in what follows.   

We begin with a basic question about the $\CRT$ theorem: how can the rotation group in Euclidean signature know something that the Lorentz group in Lorentzian signature does not?  This happens because of a curious feature of the topology of the Lorentz group.  We take the Lorentz group in $d$ dimensions to be defined as the set $O(d-1,1)$ of $d\times d$ matrices $\Lambda$ that preserve the Minkowski metric:
\be\label{lorentz}
\Lambda^\mu_{\phantom{\mu}\alpha}\Lambda^{\nu}_{\phantom{\nu}\beta}\eta_{\mu\nu}=\eta_{\alpha\beta}.
\ee
This group has four connected components, since the operations
\begin{align}\nonumber
\R:(t,x^1,x^2,\ldots x^{d-1})&\mapsto(t,-x^1,x^2,\ldots,x^{d-1})\\\nonumber
\T:(t,x^1,x^2,\ldots x^{d-1})&\mapsto(-t,x^1,x^2,\ldots,x^{d-1})\\
\R\T:(t,x^1,x^2,\ldots x^{d-1})&\mapsto(-t,-x^1,x^2,\ldots,x^{d-1})
\end{align}
cannot be continuously deformed into each other or into the identity without violating \eqref{lorentz}.  One way to see this is to note that every element of the Lorentz group must obey
\begin{align}\nonumber
\det \Lambda&=\pm 1\\
(\Lambda^0_{\phantom{0}0})^2&\geq 1,
\end{align}
so we can classify elements of the Lorentz group into four components based on the signs of $\det \Lambda$ and $\Lambda^0_{\phantom{0}0}$.  The identity, $\R$, $\T$, and $\R\T$ each live in different components under this decomposition.  This situation is to be contrasted with the Euclidean rotation group, which we define as the set $O(d)$ of $d\times d$ matrices $D$ obeying
\be
D^T D=I.
\ee
$O(d)$ has two connected components, determined by the fact that we must have $\det D=\pm 1$.  Analytic continuation from Lorentzian to Euclidean signature thus gets rid of two of the components of the Lorentz group! The way this happens is that elements of the identity and $\RT$ components of the Lorentz group both analytically continue to elements of the identity (i.e. $\det D=1$) component of $O(d)$, while elements of the $\R$ and $\T$ components both analytically continue to the $\det D=-1$ component.  Since elements of the identity component of the rotation group must be symmetries in any Euclidean relativistic field theory, some version of $\RT$ must also be a symmetry in Lorentzian signature.  

To see that the symmetry we get in Lorentzian signature is really $\CRT$ is a bit more subtle; essentially the $\C$ arises because $\RT$ must be represented by an antiunitary operator, and for any antiunitary symmetry $\Theta$ the Ward identity has the form (see equation \eqref{wardapp})
\be\label{antiunitary}
\lan\Omega|O_1\ldots O_n|\Omega\ran=\lan\Omega|(\Theta^\dagger O_n \Theta)^\dagger\ldots (\Theta^\dagger O_1 \Theta)^\dagger|\Omega\ran.
\ee
Since rotations in Euclidean signature do not introduce any complex conjugation, we need the action of $\Theta$ to take the complex conjugate of the fields in order to get something that becomes a Euclidean rotation after analytic continuation.  In more detail, in Euclidean signature we have
\begin{align}\nonumber
\lan \Omega|T O_1[\Phi]\ldots O_n[\Phi]|\Omega\ran&=\int \mathcal{D}\phi O_1[\phi]\ldots O_n[\phi]e^{-S_E[\phi]}\\\nonumber
&=\int \mathcal{D}(D_\pi\phi) O_1[D_\pi\phi]\ldots O_n[D_\pi\phi]e^{-S_E[D_\pi\phi]}\\\nonumber
&=\int \mathcal{D}\phi O_1[D_\pi\phi]\ldots O_n[D_\pi\phi]e^{-S_E[\phi]}\\
&=\lan \Omega|T O_1[D_\pi\Phi]\ldots O_n[D_\pi\Phi]|\Omega\ran.\label{CRTpath}
\end{align}
Here we adopt a notation where $\Phi$ are the fields as quantum operators and $\phi$ are the fields as path integration variables, $D_\pi$ is a counterclockwise rotation by $\pi$ in the $(x^1, t_E)$ plane, with $t_E $ the Euclidean time, and by $D_\pi\phi$ we mean the action of this rotation on the dynamical fields (which could have arbitrary spin).  In the second line we have changed variables in the path integral, while in the third line we have used that $D_\pi$ is a symmetry that leaves the measure and action invariant.  Applying a Wick rotation $t_E =it$ to both sides, which we indicate by $W$,\footnote{$W$ acts on tensor indices of the fundamental fields in the following way: for each raised $t_E $ index we supply a factor of $i$, while for each lowered $t_E $ index we supply a factor of $-i$.  It does not act on spinor indices, but it does act on the $\gamma$-matrices by sending $\gamma^{t_E }\to i\gamma^t$.} and for simplicity assuming that the operators $O_1,\ldots,O_n$ are time-ordered, we get
\be
\lan\Omega|WO_1[\Phi]\ldots WO_n[\Phi]|\Omega\ran=(-1)^{\left(F_{O_1}+\ldots F_{O_n}\right)/2}\lan\Omega|WO_n[D_\pi\Phi]\ldots WO_1[D_\pi\Phi]|\Omega\ran.
\ee
Here $F_{O_i}$ is zero if $O_i$ is bosonic and one if it is fermionic, and the sign factor arises from the time-ordering  of fermionic operators.\footnote{Note that the correlation function will be zero unless the total number of fermions $F=F_{O_1}+\ldots + F_{O_n}$ is even.  Reversing the order of $F$ fermions (with $F$ even) gives $(-1)^{F(F-1)/2}=(-1)^{F/2}$.}  This has the form \eqref{antiunitary} provided that we define an antiunitary operator $\Theta_{\CRT}$ that acts on all operators as
\be\label{CRTdef}
\Theta_{\CRT}^\dagger WO[\Phi]\Theta_{\CRT}=i^{F_O}\left(WO[D_\pi\Phi]\right)^\dagger.
\ee
We take this to be the definition of $\CRT$ symmetry.  We emphasize again that this definition does not require separate definitions of $\C$, $\R$, and $\T$.  More concretely, given a fundamental field which in Euclidean signature transforms as
\be
D_\pi \phi^a(x_E)=D_E(\R\T)^a_{\phantom{a}b}\phi^b(\R\T x_E)
\ee
in the path integral, in Lorentzian signature we have the $\CRT$ transformation\footnote{It is not immediately obvious that this definition is consistent with the requirement that the $\CRT$ transformation of the dagger of a field is the dagger of the $\CRT$ transformation of the field.  This follows from the fact that in Euclidean signature $\phi^*$ transforms under $\RT$ by the transpose of of $D_E(\R\T)$ and the unitarity of $D_E(\R\T)$.}
\be\label{CRTL}
\Theta_{\CRT}^\dagger \Phi^a(x)\Theta_{\CRT}=i^{F_\phi}\left(D_E(\R\T)^{a}_{\phantom{a}b}\right)^*\Phi^b(\R\T x)^\dagger.  
\ee
In particular the action of $\CRT$ on complex scalar, vector, and spinor fields is given by\footnote{Here $\Psi^*$ means that we take the Hilbert space adjoint of $\Psi$ but do not take the transpose of its Dirac indices.  We explain our conventions for spinors and $\gamma$ matrices in section \ref{fermsec}.}
\begin{align}\nonumber
\Theta_{\CRT}^\dagger\Phi(x)\Theta_{\CRT}&=\Phi(\RT x)^\dagger\\\nonumber
\Theta_{\CRT}^\dagger V^\mu(x)\Theta_{\CRT}&=(\RT)^{\mu}_{\phantom{\mu}\nu}V^\nu(\RT x)^\dagger\\
\Theta_{\CRT}^\dagger\Psi(x)\Theta_{\CRT}&=\gamma^{0*}\gamma^{1*}\Psi^*(\RT x).
\end{align}
There is quite a bit of confusion in the literature about whether $\CRT$ should square to $1$ or $(-1)^F$, from \eqref{CRTL} we have
\begin{align}\nonumber
\Theta_{\CRT}^\dagger\Theta_{\CRT}^\dagger\Phi^a(x)\Theta_{\CRT}\Theta_{\CRT}&=i^{-2F_\phi}D_E(\R\T)^a_{\phantom{a}b}D_E(\R\T)^b_{\phantom{a}c}\Phi^c(x)\\
&=\Phi^a(x),
\end{align}
where in the second line we have used the spin-statistics relation that $D_E(\R\T)$ squares to fermion parity $(-1)^{F_\phi}$.  Thus we see that in general we have
\be
\Theta_{\CRT}^2=1.
\ee 
Turning this into a statement about $\C\mathcal{P}\T$ in $3+1$ dimensions by combining it with a spatial rotation, we have
\be
\Theta_{\C\mathcal{P}\T}^2=(-1)^F \qquad\qquad (d=4).
\ee
Many textbooks get this wrong, one which gets it right is \cite{Haag:1992hx}.\footnote{We thank Zohar Komargodski for extensive discussion of the issue, \cite{Hason:2020yqf} is another place which gets it right (see also \cite{Wan:2023nqe}).}  Most discussions of $\CRT$, for example that in \cite{Weinberg:1995mt}, implicitly assume continuous fermion number symmetry and then define a $\CRT$ transformation which differs from \eqref{CRTL} by a factor of $-i$ on spinors and $i$ on conjugate spinors.  That version of $\CRT$ is not consistent with imposing a Majorana constraint however.  For example in $1+1$ dimensions it would lead to a Majorana-Weyl $\CRT$ transformation $\Psi'(x)=i\Psi(-x)$, which not consistent with $\Psi$ being hermitian.  This version of $\CRT$ thus can be broken e.g. by a Majorana neutrino mass, while the version \eqref{CRTL} is unbreakable.

The appearance of $\C$ from Wick rotation here applies for any time-reversing symmetry in Euclidean signature.  From this point of view the traditional definition of $\T$ in Lorentzian signature is not so natural: given a theory with $\R$ symmetry we can always rotate $\R$ by $\pi/2$ in Euclidean signature to get a time-reversing symmetry, but after analytic continuation to Lorentzian signature what we end up with is $\C\T$ instead of $\T$.  This is why in Lorentzian signature the $\CRT$ theorem ensures that $\R$ is a symmetry if and only if $\C\T$ is a symmetry.

\bfig
\includegraphics[height=4cm]{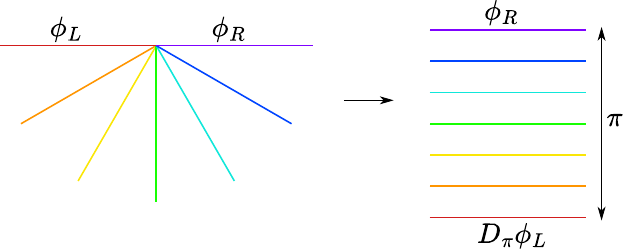}
\caption{Deriving the Rindler representation of the ground state of a relativistic field theory.}\label{rindlerfig}
\efig
An application of $\CRT$ that we will need below is its appearance in the Rindler description of the vacuum \cite{Bisognano:1976za}.  Before presenting this we first introduce some notation: given a complete commuting (or anticommuting) set of the dynamical fields at $t=0$, which we'll write as $\Phi$, we can introduce field eigenstates $|\phi\ran$ obeying
\be\label{basisdef1}
\Phi(\vec{x})|\phi\ran=\phi(\vec{x})|\phi\ran.
\ee
In bosonic theories we have $[\Phi,\Phi^\dagger]=0$, in which case these states are also eigenvectors of $\Phi^\dagger(\vec{x})$ with eigenvalue $\phi^*(\vec{x})$. Fermionic fields usually don't anticommute with their adjoints however, for example the canonical conjugate of a Dirac fermion $\Psi$ is $i\Psi^\dagger$.  We therefore introduce a separate notation for eigenstates of $\Phi^\dagger$:
\be\label{basisdef2}
\Phi^\dagger(\vec{x})|\wt{\phi}\ran=\phi^*(\vec{x})|\wt{\phi}\ran.
\ee
The dual of $|\wt{\phi}\ran$ is a left eigenstate of $\Phi$:
\be
\lan\wt{\phi}|\Phi(\vec{x})=\lan\wt{\phi}|\phi(\vec{x}).
\ee
In quantum field theory what the path integral with Dirichlet future/past boundary conditions on $\Phi$ computes are transition amplitudes of the form $\lan \wt{\phi}_f|e^{-i Ht}|\phi_i\ran$, see for example chapter 9 of \cite{Weinberg:1995mt}.  The action of $\Theta_{\CRT}$ in this basis is given by
\be\label{CRTeigen}
\Theta_{\CRT}|\phi\ran=e^{i\alpha}|\wt{\phi^{\prime*}}\ran,
\ee
as we can check by noting that
\begin{align}\nonumber
\Phi^\dagger(\vec{x})\Theta_{\CRT}|\phi\ran&=\Theta_{\CRT}\Phi^{\prime \dagger}(\vec{x})|\phi\ran\\\nonumber
&=\Theta_{\CRT}\phi^{\prime*}(\vec{x})|\phi\ran\\
&=\phi^\prime(x)\Theta_{\CRT}|\phi\ran.
\end{align}
The phase $e^{i\alpha}$ must be independent of $\phi$ in order for $\CRT$ to act correctly on the canonical momenta (as $\Pi'=-i\frac{\delta}{\delta \phi'}$), and can thus be absorbed into a redefinition of $\Theta_{\CRT}$ (this doesn't affect our conclusion that $\Theta_{\CRT}^2=1$ since $(e^{-i\alpha}\Theta_{\CRT})^2=\Theta_{\CRT}^2$ by antilinearity), so from now on we set $\alpha=0$.\footnote{One might worry that this rephasing causes the action of $\Theta_{\CRT}$ on the vacuum to include a nontrivial phase, but we can always remove such a phase by rephasing the vacuum. }

We will also need to know how these field eigenstates decompose under tensor products.  Namely we have
\begin{align}\nonumber
|\phi_A\phi_B\ran_{AB}&=|\phi_A\ran_A\otimes |\phi_B\ran_B\\
\lan \wt{\phi_A\phi_B}|_{AB}&=\lan\wt{\phi_B}|_B\otimes \lan\wt{\phi_A}|_A.
\end{align}
The reversed ordering in the second line arises because in a fermionic system the state $|1\ran$ is Grassmann odd while the state $|0\ran$ is Grassman even, so to have $\lan 00|00\ran=\lan 11|11\ran=1$ we need the dual to reverse the order of the tensor product.  The right eigenstates $|\phi\ran$ are always Grassmann even, but the left eigenstates $\lan \wt{\phi}|$ have Grassmann weight $(-1)^M$ where $M$ is the number of complex fermions (see for example equation 9.5.18 of \cite{Weinberg:1995mt}).  From \eqref{CRTeigen} we thus see that $\Theta_{\CRT}$ has Grassmann weight $(-1)^M$.  We will explain these weights in more detail in an upcoming paper.\footnote{Even in bosonic theories it is arguably more natural to define the dual of a tensor product to reverse the order, as then we can contract the bras and kets ``from the inside out'': $\big(\lan \chi'|\otimes \lan \phi'|\big)\big(|\phi\ran\otimes |\chi\ran\big)=\lan \chi'|\big(\lan\phi'|\phi\ran\big)|\chi\ran=\lan\chi'|\chi\ran\lan\phi'|\phi\ran$.}  We will also assume for the sake of convenience that the theory we are considering has a well-defined square root of fermion parity that we will call $i^F$.  In practice what this means is that we can group the fermions into complex fields in such a way that $\Psi'=i\Psi$ and $\Psi^{\dagger \prime}=-i\Psi^\dagger$ for all fermion fields is a symmetry.  This will of course be true if there is a continuous fermion number symmetry $\Psi'=e^{i\theta}\Psi$, as in quantum electrodynamics, but we won't need this stronger assumption.  We will discuss the general case (without $i^F$ symmetry) in our upcoming paper.    

Turning now to the Rindler description of the vacuum in theories with $i^F$ symmetry, if we split the Hilbert space at $t=0$ into $\mathcal{H}_L\otimes\mathcal{H}_R$ then from the Euclidean path integral we have (see figure \ref{rindlerfig})
\begin{align}\nonumber
\lan\wt{\phi_L\phi_R}|\Omega\ran&\propto\lan\wt{\phi_R}|e^{-\pi K_R}|D_\pi \phi_L\ran\\\label{rindler}
&=\sum_n e^{-\pi \omega_n}\lan\wt{\phi_R}|n\ran\lan n|D_\pi\phi_L\ran,
\end{align}
where 
\be
K_R=\int_{x^1>0}d^{d-1}x\, x^1 T_{tt}
\ee
is the right boost operator and $|n\ran$ is a complete eigenbasis for $K_R$ with eigenvalue $\omega_n$.\footnote{In continuum field theory this argument commits a number of petty mathematical crimes, starting with the illegal split of the Hilbert space into a tensor product and continuing with the treatment of $K_R$ as a valid operator.  To justify it rigorously one needs to either regulate the theory or adopt a more formal approach as in \cite{Bisognano:1976za}.  We will not concern ourselves with such things here.}
 To turn \eqref{rindler} into an expression for the ground state, we first introduce an antiunitary ``restricted  $\CRT$'' operator $\Theta_{\CRT}^R:\mathcal{H}_R\to\mathcal{H}_L$ defined by
\be
\Theta_{\CRT}^R|\phi_R\ran=|\wt{\phi_L^{\prime *}}\ran=|\wt{i^{-F_\phi}D_\pi \phi_R}\ran.
\ee
Substituting $\phi_R\to D_\pi\phi_L$ we have
\begin{align}\nonumber
\Theta_{\CRT}^R|D_\pi\phi_L\ran&=|\wt{i^{F_\phi}\phi_L}\ran\\
&=i^{F_L}|\wt{\phi_L}\ran,
\end{align}
where $i^{F_L}$ implements $i^F$ symmetry on $\mathcal{H}_L$.  Defining
\be
\Theta_R=i^{-F_L}\Theta_\CRT^R,
\ee  
we thus have
\be\label{rindlerreq}
\Theta_R|D_\pi\phi_L\ran=|\wt{\phi_L}\ran.
\ee
From \eqref{rindler} we thus have
\begin{align}\nonumber
\lan \wt{\phi_L\phi_R}|\Omega\ran&=\left(\lan \wt{\phi}_R| \lan \wt{\phi_L}|\right)|\Omega\ran\\\nonumber
&\propto \sum_n e^{-\pi \omega_n}\lan \wt{\phi}_R|n\ran_R\lan \Theta_R D_\pi \phi_L|\Theta_R n\ran_L\\\nonumber
&=\sum_n e^{-\pi \omega_n}\lan \wt{\phi}_R|n\ran_R\lan \wt{\phi_L}|\Theta n\ran_L\\
&=\lan \wt{\phi}_R| \lan \wt{\phi_L}|\sum_n (-1)^{F_n}e^{-\pi \omega_n}|\Theta_R n\ran_L|n\ran_R,\label{rindlerdev}
\end{align}
where the factor of $(-1)^{F_n}$ in the last line comes from exchanging the weight $F_n$ object $\lan \wt{\phi_L}|\Theta_R n\ran_L$ and the weight $F_n$ object $|n\ran_R$ (note that $\lan \wt{\phi_L}|$ and $\Theta_R$ both have weight $M/2$, where $M/2$ is the (divergent) number of complex fermions in each wedge).  Comparing the first and last lines of \eqref{rindlerdev}, we see that up to normalization the ground state is given by
\be\label{Rindlerstate}
|\Omega\ran\propto \sum_n e^{-\pi \omega_n} i^{F_L}|\Theta_{\CRT}^Rn\ran_L\otimes |n\ran_R.
\ee
We have not seen a general discussion of the factor of $i^{F_L}$ appearing here in the previous literature, perhaps because in theories with $i^F$ symmetry it can be absorbed into a redefinition of $\CRT$ symmetry.  It has however appeared in some SYK calculations \cite{Maldacena:2018lmt,Garcia-Garcia:2019poj,Nosaka:2022lqd}, where it was motivated by the idea that the infinite-temperature thermofield double state should be annihilated by a fermionic lowering operator.\footnote{SYK is an example which doesn't have $i^F$ symmetry, but \eqref{Rindlerstate} is still correct in such theories with an appropriate definition of $i^{F_L}$, as we will explain in our upcoming paper.}  

In what follows we will also be interested in the vacuum wave function in a basis where we fix $\phi_L^*$ instead of $\phi_L$, so we'll note now that from figure \eqref{rindlerfig} have
\be
\lan \phi_L\wt{\phi_R}|\Omega\ran\propto \lan \wt{\phi_R}|e^{-\pi K_R}|\wt{D_\pi\phi_L}\ran.
\ee
Defining
\be
\Theta_L=i^{-F_R}\Theta^L_{\CRT}
\ee
with
\be
\Theta^L_{\CRT}|\phi_L\ran=|\wt{\phi_L^{\prime*}}\ran,
\ee
we again have
\be
\Theta_L|D_\pi\phi_R\ran=|\wt{\phi_R}\ran.
\ee
Substituting $\phi_R=D_\pi \phi_L$ we thus have
\be
|\wt{D_\pi\phi_L}\ran=\Theta_L(-1)^{F_L}|\phi_L\ran=i^{F_R}\Theta^L_{\CRT}|\phi_L\ran,
\ee
so the ground state wave function with mixed boundary conditions can be written as
\be
\lan \phi_L\wt{\phi_R}|\Omega\ran\propto \lan \wt{\phi_R}|e^{-\pi K_R}i^{F_R}\Theta_\CRT^L|\phi_L\ran.\label{mixedrindler}
\ee


\section{Background gauge fields for spacetime symmetries in field theory}\label{backsec}
We now begin the process of gauging spacetime inversions.  As we reviewed in section \ref{intsec}, the first step of gauging is turning on a background gauge field.  As a warmup we'll briefly discuss background gauge fields for spatial translations, and then move on to discuss background gauge fields for both $\R$ and $\T$.  For concreteness we work in free scalar field theory, mostly in $1+1$ dimensions, but our general conclusions do not depend on this.

\subsection{Background fields for spatial translation}
\bfig
\includegraphics[height=4cm]{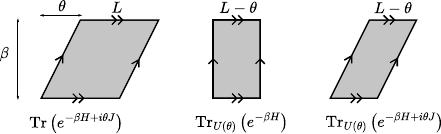}
\caption{Background fields for a spatial rotation.}\label{Jfig}
\efig
Consider again a free scalar field in $1+1$ dimensions with a compact target space radius $R_\phi$, quantized on a spatial circle of circumference $L$. There is a global spatial rotation symmetry, which acts on the fields as
\be
e^{-i\theta J}\Phi(t,x)e^{i\theta J}=\Phi\left(t,x-\frac{\theta}{2\pi}L\right).
\ee
In order to turn on a background gauge field for this symmetry, we should insert
\be
U(\theta)=e^{i\theta J}
\ee
on some cycle(s) in spacetime.  Perhaps the most straightforward thing to do is to wrap $U(\theta)$ on the spatial circle.  Computing the thermal trace with such an insertion gives
\be
Z_1=\Tr\left(e^{-\beta H+i\theta J}\right),
\ee    
which is precisely the partition function we studied in section \ref{intexsec}.  Thus we see that we can interpret $\theta$ in \eqref{Z1bos} as introducing a background gauge field for spatial rotation.  Geometrically we can view this as tilting the torus, as shown in figure \ref{Jfig}.

More nontrivial is to extend $U(\theta)$ in the time direction, which creates a twisted sector with spatial holonomy for spatial rotation.  Indeed as we move $\Phi$ around the spatial circle, the presence of $U(\theta)$ extended in time changes the boundary conditions of the scalar to
\be
\Phi\left(t,x+L\left(1-\frac{\theta}{2\pi}\right)\right)=\Phi(t,x)+2\pi R_\phi m,
\ee
where $m$ again is the winding number of the scalar about the spatial circle.  In other words the insertion of $U(\theta)$ extended in the time direction changes the spatial periodicity from $L$ to $L\left(1-\frac{\theta}{2\pi}\right)$.  We can also include $U(\theta)$ in both the spatial and temporal directions, which both shrinks the spatial radius and tilts the torus (see figure \ref{Jfig}).

In higher dimensions we can also consider twisted sectors where going around one spatial direction introduces a rotation in a different spatial direction:
\be
\Phi(t,x+L_x,y)=\Phi\left(t,x,y-\frac{\theta}{2\pi}L_y\right).
\ee
This introduces tilting into the spatial geometry.

The key point illustrated by these examples is the following: \textbf{a background gauge field for a spacetime symmetry is a change in the geometry of spacetime.}  Of course for Poincare symmetries such as translations this is a very standard observation: the spacetime metric is the background gauge field for Poincare symmetry.  Indeed by a linear transformation of the coordinates we can convert a change in spatial periodicity and/or a twist in the thermal trace to a change in the background metric; this is one way of seeing how summing over metrics automatically gauges translation symmetry.

\subsection{Background fields for $\R$}\label{Rsubsec}

\bfig
\includegraphics[height=5cm]{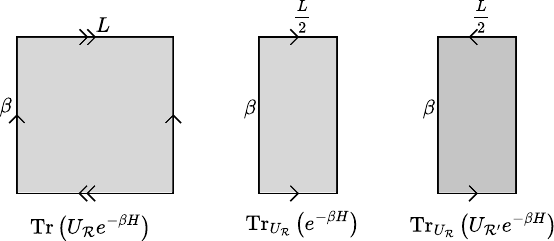}
\caption{Background fields for $\R$ symmetry in $1+1$ dimensions.  Inserting $U_\R$ wrapping a spatial circle results in the partition function on the Klein bottle, while inserting $U_\R$ on a temporal circle results in the partition function on a spatial interval of length $L/2$ with Neumann boundary conditions at both ends. Inserting $U_\R$ on the temporal circle and $U_{\R'}$ on the spatial circle gives a M\"obius strip.}\label{Rbackfig}
\efig
We now consider background gauge fields for spatial reflection symmetry.\footnote{The gauging of $\R$ was recently discussed from a more phenomenological point of view in \cite{McNamara:2022lrw}, whose treatment is compatible with our approach here.}  For simplicity we will from now on take the radius $R_\phi$ of our scalar to infinity to avoid further discussion of winding modes (this also is a better model for the analogous constructions in higher dimensions). We'll take $U_\R$ to act as
\be
U_\R^\dagger \Phi(t,x)U_\R=\Phi(t,L-x).
\ee
Inserting $U_\R$ on a spatial slice in the thermal trace results in a path integral over Euclidean fields obeying
\be
\phi(t_E +\beta,x)=\phi(t_E ,L-x),
\ee
which we can interpret as computing the Euclidean path integral on a Klein bottle (see figure \ref{Rbackfig}).  We thus again see that a background gauge field for a spacetime symmetry changes the geometry (and in this case the topology) of spacetime.   

We can also insert $U_\R$ extending in the time direction.  This introduces an identification 
\be
\phi(t,x+L)=\phi(t,-x),
\ee
or equivalently
\be
\phi(t,x)=\phi(t,L-x),
\ee
which we can interpret this as changing the spatial topology from a circle of circumference $L$ to an interval of length $L/2$.  The boundary conditions at the endpoints are Neumann boundary conditions, since
\begin{align}\nonumber
\partial_x \phi(t,L/2)&=\lim_{\epsilon\to 0}\frac{\phi(t,L/2+\epsilon)-\phi(t,L/2-\epsilon)}{2\epsilon}\\
&=0
\end{align}
and
\begin{align}\nonumber
\partial_x\phi(t,0)&=\lim_{\epsilon\to 0}\frac{\phi(t,\epsilon)-\phi(t,-\epsilon)}{2\epsilon}\\\nonumber
&=\lim_{\epsilon\to 0}\frac{\phi(t,\epsilon)-\phi(t,L-\epsilon)}{2\epsilon}\\
&=0.
\end{align}
Thus we see that a background gauge field for a spacetime symmetry can introduce a boundary of spacetime!  In string theory this has an interesting consequence: it says that in unoriented string theory open strings arise as twisted sectors from gauging worldsheet parity \cite{Sagnotti:1987tw,Pradisi:1988xd}.\footnote{There seems to have been some skepticism about this claim in the literature, see e.g. the end of lecture one in \cite{Polchinski:1996fm} or the end of chapter 8 in \cite{Polchinski:1998rq}.  We are in agreement with \cite{Sagnotti:1987tw,Pradisi:1988xd}: open strings are twisted states of the unoriented string in just the usual way, and not including them when worldsheet parity is gauged is a violation of locality on the worldsheet.  It would be interesting to try to develop this idea into a non-perturbative argument that in bosonic string theory the unoriented string is inconsistent if open strings are not included, in particular Emil Martinec has suggested this would be natural in the context of the matrix models of \cite{Harris:1990kc,Brezin:1990xr}.}  For example this is true in the relationship between the type IIB string and the type I string.\footnote{In the type I case we are gauging both worldsheet parity and fermion parity, and the combination gives the GSO projection of the open type I string.  One could also consider orientifolds, where one gauges a combination of worldsheet parity and target space reflection, which seems likely to give D branes of lower dimension.}  

It might seem that the natural next step is to have $U_\R$ wrapping circles in both space and time, but this is actually trivial: the spatial holonomy has removed the region with $x\in (L/2,L)$, so there is nothing left to reflect.  What we can do instead is include a reflection about $x=L/4$ via an operator $U_{\R'}$ which acts as
\be
U_{\R'}^\dagger \Phi(t,x)U_{\R'}=\Phi(t,L/2-x),
\ee
which results in the partition function on the M\"obius strip, as shown in figure \ref{Rbackfig}.  We emphasize that both this example and the Klein bottle are not orientable manifolds, so a background gauge field for $\R$ can give rise to an unoriented spacetime.

\subsection{Background fields for $\T$ part I: Hilbert space and Heisenberg operators}\label{tback1}
Life becomes more interesting once we turn on background fields for $\T$. We will start in Lorentzian signature, which after all is the place where things happen.  

\bfig
\includegraphics[height=5cm]{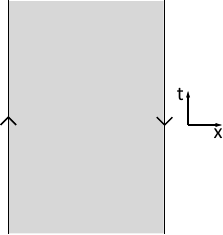}
\caption{The Lorentzian M\"obius strip: a time-unorientable Lorentzian manifold.}\label{lorentzmobfig}
\efig
Time reversal in quantum field theory is implemented by an antiunitary operator $\Theta_\T$, which acts on a real scalar field as
\be
\Theta_\T^\dagger\Phi(t,x)\Theta_\T=\Phi(-t,x).
\ee
We can certainly include a background gauge field in the untwisted sector by wrapping $\Theta_\T$ on a spatial slice, and summing over such fields restricts us to $\T$-invariant states as one would expect.  What is more mysterious is defining the twisted sectors.  The natural proposal for what it would mean to insert $\Theta_\T$ extended in the time direction for our free scalar theory would be to introduce a twisted a Hilbert space where the field obeys the boundary condition
\be\label{Tbc}
\Phi(t,x+L)=\Phi(-t,x).
\ee
We can try to interpret this geometrically as studying the theory on a ``Lorentzian M\"obius strip'' where the orientation of time changes as we move around a spatial circle, see figure \ref{lorentzmobfig}.  Here however we meet a problem: to construct a quantum scalar field theory on this geometry we would like to introduce the canonical commutation relation
\be\label{cancomm}
[\Phi(0,x),\dot{\Phi}(0,y)]=i\delta(x-y),
\ee
but how are we supposed to choose the direction of the time coordinate in $\dot{\Phi}$ given that this direction changes sign as we move around the circle?  In the past this problem has been used to argue that quantum field theory does not make sense on manifolds that are not time-orientable \cite{Kay:1992es,Friedman:1995mr}.  We think that it does make sense.  The reason is that it is ok for the canonical commutation relation to change sign discontinuously as we cross an insertion of $\Theta_\T$ in the path integral.  The ability of local correlators to detect the location of $\Theta_\T$ is no more pathological here than it was for background gauge fields of discrete internal symmetries, as we discussed in section \ref{intsec}.  The important thing is that we lose the ability to detect this location once we gauge the symmetry, and that is true here since $\dot{\Phi}$ is not $\T$-invariant and thus is not a valid observable in the gauged theory.  In this subsection and the following two we construct this quantum theory in detail and do a number of calculations to develop some intuition for life in a time non-orientable spacetime.

We begin with the classical theory of a scalar field on the Lorentzian M\"obius strip.  The boundary conditions \eqref{Tbc} constrain the set of initial data at $t=0$ to be such that $\phi(0,x)$ is periodic and $\dot{\phi}(0,x)$ is antiperiodic.  A complete set of solutions for this initial data is given by
\begin{align}\nonumber
\phi(t,x)=A_0+\sum_{n=1}^\infty\Bigg[&\left(A_n\cos\left(\omega_n x\right)+B_n\sin\left(\omega_n x\right)\right)\cos(\omega_n t)\\
&+\left(C_n \cos\left(\hat{\omega}_n x\right)+D_n\sin\left(\hat{\omega}_n x\right)\right)\sin(\hat{\omega}_nt)\Bigg],\label{phisol}
\end{align}
with
\begin{align}\nonumber
\omega_n\equiv&\frac{2\pi n}{L}\\
\hat{\omega}_n\equiv&\frac{2\pi (n-1/2)}{L}.
\end{align}
To see that this set is general enough to account for all initial data we can invert it, finding 
\begin{align}\nonumber
A_n&=\frac{2}{L}\int_0^Ldx\cos(\omega_nx)\phi(0,x)\\\nonumber
B_n&=\frac{2}{L}\int_0^Ldx\sin(\omega_nx)\phi(0,x)\\\nonumber
C_n&=\frac{2}{\hat{\omega}_nL}\int_0^Ldx\cos(\hat{\omega}_nx)\dot{\phi}(0,x)\\
D_n&=\frac{2}{\hat{\omega}_nL}\int_0^Ldx\sin(\hat{\omega}_nx)\dot{\phi}(0,x).
\end{align}
We thus have a robust phase space of classical solutions for a fairly reasonable set of initial data.  

\bfig
\includegraphics[height=5cm]{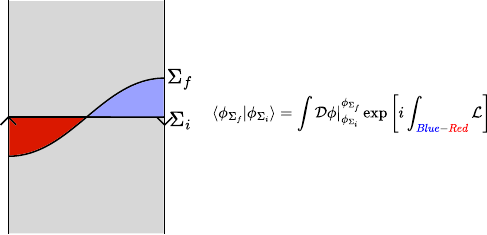}
\caption{The Lorentzian path integral for field theory on the Lorentzian M\"obius strip, computing the transition between field eigenstates on different spatial slices.  Note that many points in the geometry do not lie on any spatial slice.}\label{mobpathfig}
\efig
To canonically quantize this theory we need to impose the commutation relations \eqref{cancomm}, which we will do simply accepting the discontinuity at $x=L$ this creates for the reasons explained above.  We therefore have a Hilbert space spanned by states of the form $|\phi(x)\ran$, on which we have the algebra
\begin{align}\nonumber
[A_n,D_m]&=\frac{4i}{L^2\hat{\omega}_m}\int_0^Ldx \cos(\omega_n x)\sin(\hat{\omega}_mx)\\\nonumber
&=\frac{2i}{\pi^2((m-1/2)^2-n^2)}\\\nonumber
[B_n,C_m]&=\frac{4i}{L^2\hat{\omega}_m}\int_0^Ldx \sin(\omega_n x)\cos(\hat{\omega}_mx)\\
&=-\frac{2in}{\pi^2(m-1/2)((m-1/2)^2-n^2)},\label{commutators}
\end{align} 
with all other commutators involving $A_n\ldots D_n$ vanishing.  Here we have used the integrals
\begin{align}\nonumber
\int_0^L dx \cos(\omega_n x)\cos(\hat{\omega}_mx)&=0\\\nonumber
\int_0^L dx \sin(\omega_n x)\sin(\hat{\omega}_mx)&=0\\\nonumber
\int_0^L dx \cos(\omega_n x)\sin(\hat{\omega}_mx)&=\frac{L}{\pi}\frac{m-1/2}{(m-1/2)^2-n^2}\\
\int_0^L dx \sin(\omega_n x)\cos(\hat{\omega}_mx)&=\frac{L}{\pi}\frac{n}{n^2-(m-1/2)^2}.
\end{align}
One can confirm that this algebra is internally consistent by constructing it explicitly for a regulated version of the theory, as we explain in appendix \ref{regalg}, and the expression \eqref{phisol} gives the complete solution for the Heisenberg equations of motion.  Given the canonical commutators we can then construct a Lorentzian path integral formalism for computing the transition matrix between field eigenstates on different spatial slices, as shown in figure \ref{mobpathfig}.

\subsection{Background fields for $\T$ part II: correlation functions}
\bfig
\includegraphics[height=6cm]{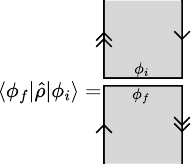}
\caption{Preparing the M\"obius pseudostate $\hat{\rho}$ using a Euclidean path integral.  $\hat{\rho}$ is not positive definite so we do not call it a state, but we can still use it to compute correlation functions.}\label{mstatefig}
\efig
Unlike the examples we have been considering until now, there is no particularly natural pure state to consider in the Hilbert space of the Lorentzian M\"obius strip.  A ground state would require time-translation invariance, which we don't have, and there is no  natural Euclidean preparation of a pure state since the boundary conditions identify boundaries to the future and past of $t=0$. The best we can do is define a mixed ``state'' $\hat{\rho}$ by cutting the Euclidean path integral on the M\"{o}bius strip, as shown in figure \ref{mstatefig}.  We say ``state'' because the operator prepared in this way is not actually positive, or even hermitian, so from now on we will refer to it as the ``M\"obius pseudostate''.  This lack of a natural state may seem concerning, and indeed has been a source of worry in previous literature \cite{Kay:1992es,Friedman:1995mr}, but we emphasize that the Hilbert space constructed in the previous section is perfectly conventional.  We are not viewing the M\"obius pseudostate as a physical configuration of the theory, rather it is a way to define a fairly natural set of ``computables'' that we can use to probe the theory given the lack of a natural state.  In section \ref{adssec} we will see that such computables are also natural to consider in the context of holography. We will describe the M\"obius pseudostate more explicitly for any 2D CFT in the next subsection, but first we compute some of its correlation functions in the free scalar theory and check that they are compatible with our canonical algebra \eqref{cancomm}.

\bfig
\includegraphics[height=6cm]{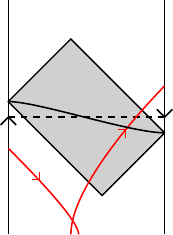}
\caption{A self-intersecting timelike curve in the Lorentzian M\"obius strip, with the $t=0$ slice shown as a dashed line.  Note however that the intersection does not happen within the grey diamond sitting on the black spatial slice: within this diamond we should expect reasonable causal behavior.}\label{causalmob}
\efig
Before getting into the details of the correlators, it is perhaps useful to first advertise the basic lesson.  This is that for any spatial slice, meaning a smooth hypersurface whose normal vector is everywhere timelike, the dynamics and causal structure of the correlation functions are conventional in a certain causal diamond associated to the slice.  Formally this diamond is defined as as the domain of dependence of the set consisting of the spatial slice minus its intersection with the support of the time reversal operator.\footnote{Note that the location of the symmetry operator therefore plays an important role in determining the region where the correlation functions are causal.  This will no longer be the case once we gauge $\T$.} On the other hand once we consider correlation functions which do not have all points in any such diamond there can be unusual causal features.  Intuitively these features arise from the existence of self-intersecting timelike curves (SITCs), with an example being shown in figure \ref{causalmob}.  The Lorentzian M\"obius strip does not have genuine closed timelike curves (CTCs), but SITCs are already sufficient to lead to causal pathologies such as meeting a future version of yourself.  Indeed in a time-orientable spacetime one can always deform an SITC into a CTC, which presumably is why most discussions of causal pathologies in the relativity literature focus on CTCs (see for example \cite{Hawking:1991nk}, where time-orientability is assumed), but in spacetimes that are not time orientable SITCs are a better bellwether.  In a gravitational context one can hope that any timelike curve must pass behind an event horizon before intersecting itself, and indeed our expectation is that the grey diamond in figure \ref{causalmob} is a good model for the spacetime region that is exterior to all horizons (sometimes called the domain of outer communication).  We return to this point in section \ref{adssec}.

\bfig
\includegraphics[height=6cm]{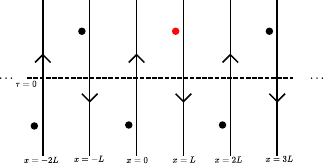}
\caption{Images for computing the Euclidean correlation function on the M\"{o}bius strip.  The genuine operator in the fundamental domain is the red dot, while the black dots are the image operators.}\label{imagesfig}
\efig
Moving now to the correlation functions, it is convenient to use complex Euclidean coordinates $z=x+it_E $ with the M\"{o}bius boundary conditions implying that
\be\label{mobiusbc}
\phi(z+L,\ol{z}+L)=\phi(\ol{z},z).
\ee
The Euclidean Green function is naively obtained (see figure \ref{imagesfig}) by a sum over images:
\be\label{naivemobius}
G(z_1,\ol{z}_1;z_2,\ol{z}_2)\equiv\lan\phi(z_2,\ol{z}_2)\phi(z_1,\ol{z}_1)\ran= -\frac{1}{4\pi}\sum_{n=-\infty}^\infty\Big[\log|z_2-z_1+2 n L|^2+\log|z_2-\ol{z}_1+(2n+1)L|^2\Big].
\ee 
This sum formally obeys
\begin{align}\nonumber
G(z_1+L,\ol{z}_1+L;z_2,\ol{z}_2)&=G(\ol{z}_1,z_1;z_2,\ol{z}_2)\\
G(z_1,\ol{z}_1;z_2+L,\ol{z}_2+L)&=G(z_1,\ol{z}_1;\ol{z}_2,z_2),
\end{align}
and thus is consistent with \eqref{mobiusbc}.  Unfortunately however the sum is divergent.  This divergence has nothing to do with the non-orientable nature of the strip, it is just a consequence of the noncompact zero mode of the scalar and the same would be true for the Green function on a torus.  To resolve this we should instead consider derivatives of $G$, for which the sum is convergent.  These are the correlation functions of the primary operators $\partial_z\phi$ and $\partial_{\ol{z}}\phi$.  Explicitly we have
\begin{align}\nonumber
\partial_{z_1}\partial_{z_2}G&=-\frac{1}{4\pi}\sum_{n=-\infty}^\infty\frac{1}{(z_2-z_1+2nL)^2}=-\frac{\pi}{16L^2\sin^2\left(\frac{\pi}{2L}\left(z_2-z_1\right)\right)}\\\nonumber
\partial_{\ol{z}_1}\partial_{\ol{z}_2}G&=-\frac{1}{4\pi}\sum_{n=-\infty}^\infty\frac{1}{(\ol{z}_2-\ol{z}_1+2nL)^2}=-\frac{\pi}{16L^2\sin^2\left(\frac{\pi}{2L}\left(\ol{z}_2-\ol{z}_1\right)\right)}\\\nonumber
\partial_{z_1}\partial_{\ol{z}_2}G&=-\frac{1}{4\pi}\sum_{n=-\infty}^\infty\frac{1}{(\ol{z}_2-z_1+(2n+1)L)^2}=-\frac{\pi}{16L^2\cos^2\left(\frac{\pi}{2L}\left(\ol{z}_2-z_1\right)\right)}\\
\partial_{\ol{z}_1}\partial_{z_2}G&=-\frac{1}{4\pi}\sum_{n=-\infty}^\infty\frac{1}{(z_2-\ol{z}_1+(2n+1)L)^2}=-\frac{\pi}{16L^2\cos^2\left(\frac{\pi}{2L}\left(z_2-\ol{z}_1\right)\right)}.
\end{align}

To compare to the Hilbert space formalism, we should compute the Wightman functions in Lorentzian signature.  This is done by using the analytic continuations
\begin{align}\nonumber
z_1&=-(x_1^--i\epsilon_1)\\\nonumber
\ol{z}_1&=x_1^+-i\epsilon_1\\\nonumber
z_2&=-(x_2^--i\epsilon_2)\\
\ol{z}_2&=x_2^+-i\epsilon_1,
\end{align}
where
\be
x^{\pm}=t\pm x
\ee 
and we take $\epsilon_2>\epsilon_1>0$ since we want to have $\phi(x_2^+,x_2^-)$ appear to the left of $\phi(x_1^+,x_1^-)$.  Thus we have
\begin{align}\nonumber
\partial_{x_1^-}\partial_{x_2^-}G&=-\frac{\pi}{16L^2\sin^2\left(\frac{\pi}{2L}\left(x_2^--x_1^--i\epsilon\right)\right)}\\\nonumber
\partial_{x_1^+}\partial_{x_2^+}G&=-\frac{\pi}{16L^2\sin^2\left(\frac{\pi}{2L}\left(x_2^+-x_1^+-i\epsilon\right)\right)}\\\nonumber
\partial_{x_1^-}\partial_{x_2^+}G&=\frac{\pi}{16L^2\cos^2\left(\frac{\pi}{2L}\left(x_2^++x_1^--i\epsilon\right)\right)}\\
\partial_{x_1^+}\partial_{x_2^-}G&=\frac{\pi}{16L^2\cos^2\left(\frac{\pi}{2L}\left(x_2^-+x_1^+-i\epsilon\right)\right)},\label{wightman}
\end{align}
with $\epsilon>0$.  We can think of these four correlation functions as giving the right-moving response to a right-moving source, the left-moving response to a left-moving source, the left-moving response to a right-moving source, and the right-moving response to a left-moving source.  With the other operator ordering $\epsilon$ changes sign in the first two lines but not in the second two.  

\bfig
\includegraphics[height=5cm]{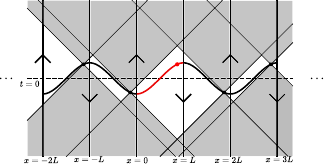}
\caption{Commutativity of operators at spacelike separation on a spatial slice of the Lorentzian M\"{o}bius strip: for any operator on a spatial slice of the fundamental domain, shaded red, the images of that operator (shown as black dots) are spacelike with respect to the entire red slice.}\label{images2fig}
\efig
The first two lines of \eqref{wightman} are the same correlators we would have on a Lorentzian cylinder of diameter $2L$, while the second two incorporate the fact that a right/left mover can go around the M\"{o}bius strip once and turn into a left/right mover.  We can think of the singularities in \eqref{wightman} as arising whenever one of the operators crosses the future/past light cone of the other operator \textit{or any of its images}.  The important point is that whenever an operator lies on a spatial slice of the fundamental domain, then the rest of that slice is always spacelike from all of the images of that operator (see figure \ref{images2fig}).  This ensures that the algebra of operators on any spatial slice is compatible with the canonical commutation relations, and thus that the correlation functions of operators in the causal diamond sitting over that slice (as in figure \ref{causalmob}) have standard causal properties.\footnote{Netta Engelhardt points out that this example suggests a general criterion for being able to construct a quantum field theory on a time non-orientable spacetime $M$: it should have an acausal hypersurface $\Sigma$ with the property that it splits the time-oriented double cover of $M$ into two disjoint pieces.  She also recommends we take the double cover to be \textit{stably causal}, in the sense that there exists a (highly non-unique) ``time function'' whose gradient gives a nowhere-vanishing timelike vector.  The existence of such a hypersurface with a stably causal double cover is a rather mild regularity condition on the causal structure of spacetime, and it would be interesting to understand its general consequences.}  

\subsection{Background fields for $\T$ part III: identifying the M\"obius pseudostate}

We now give a more formal discussion of the Mobius pseudostate in general 2D CFTs with a time-reversal symmetry $\T$.  For convenience we will restrict to the case where this symmetry squares to the identity.  We first need to confront a subtletly that we have so far avoided in this section: as explained at the end of section \ref{topsec}, $\T$ symmetry in Lorentzian signature continues to $\C\T$ symmetry in Euclidean signature.  We avoided this because we considered a real scalar.  Since we would now like to consider a general CFT, we need to decide whether we are constructing a nontrivial background for $\T$ or $\C\T$.  If we do the latter then in the Euclidean picture we have the standard M\"obius strip, while if we do the former then in Euclidean signature we should define the pseudostate to include a complex conjugation in addition to a reflection of time as we go around the spatial circle.  Either choice is reasonable and can be analyzed, our choice will be to stick to $\T$ in Lorentzian signature, in which case the M\"obius gluing in Euclidean signature includes a complex conjugation.  For example for a complex scalar we would have 
\be
\phi(t_E,x+L)=\phi^*(-t_E,x),
\ee  
where $\phi^*(t_E,x)=\phi(-t_E,x)^\dagger$ is the Euclidean adjoint.  In the general case we will take $\T$ to act on the Euclidean fields as 
\be
\phi'(x)=D_\mathcal{T}^*\phi^*(\mathcal{T}x),
\ee
where $D_\T$ is some matrix acting on the fundamental fields.  In Lorentzian signature this induces an antiunitary symmetry $\Theta_\T$ such that
\be
\Theta_{\mathcal{T}}^\dagger \Phi(x)\Theta_{\mathcal{T}}=i^{F_\phi}D_\mathcal{T}\Phi(\mathcal{T}x).
\ee
The factor of $i^{F_\phi}$ here arises from fermion re-ordering in the same way it did for $\CRT$. Our requirement that $\Theta_\T$ squares to one tells us that
\be
D_\T D_\T^*=1.
\ee

\bfig
\includegraphics[height=5cm]{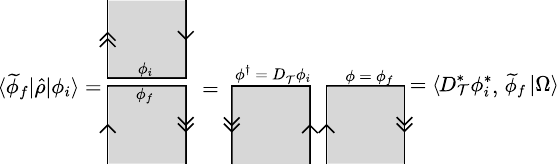}
\caption{Relationship between the Mobius pseudostate with circumference $L$ and the circle ground state with circumference $2L$.  The complex conjugation in the Euclidean gluing is removed when we reflect the upper half down, so the gluings on the right do not have any complex conjugations.  On the other hand the conjugation means that the ``final state'' on the left side is for $\Phi^\dagger$ instead of $\Phi$, so in theories where the distinction matters (i.e. theories with fermions) we are computing the ground state wave function in the ``tilde'' basis on the right side but the un-tilded basis on the left side (see equations \eqref{basisdef1}, \eqref{basisdef2} for the explanation of this distinction).}\label{mobcylfig}
\efig
The key point is that we can relate the matrix elements of the Mobius pseudostate on a circle of circumference $L$ to the ground state wave function of the same CFT studied on a circle of circumference $2L$.  The relationship is shown in figure \ref{mobcylfig}.   The idea is that if we denote the fields with $t>0$ as $\phi_+$ and the fields with $t<0$ by $\phi_-$, the gluing in the Mobius pseudostate imposes the boundary conditions
\begin{align}\nonumber
\phi_+(t,L)&=D_\T^*\phi_-^*(-t,0)\qquad (t>0)\\
\phi_-(t,L)&=D_\T^*\phi_+(-t,0)\qquad (t<0).
\end{align}
By introducing the field redefinition
\be
\hat{\phi}_+(t,x)=D_T^*\phi_+^*(-t,x+L),
\ee
these boundary conditions become
\begin{align}\nonumber
\hat{\phi}_+(t,0)&=\phi_-(t,0)\\
\hat{\phi}_+(t,-L)&=\phi_-(t,L).
\end{align}
These are of course the standard boundary conditions for studying the theory on the Euclidean half-cylinder with circle circumference $2L$.  The boundary conditions at $t=0$ are
\begin{align}\nonumber
\phi_-(0,x)&=\phi_f(x) \hspace{.5cm}\qquad\qquad (0<x<L)\\
\hat{\phi}_+^*(0,x)&=D_\T\phi_i(x+L)\qquad (-L<x<0).\label{t0conds}
\end{align}
Thus we just need to understand the ground state entanglement between the two halves of a spatial circle in 2D CFT.  This problem is conformally equivalent to the Rindler decomposition of the ground state on an infinite line \cite{Cardy:2016fqc}, which we saw in equation \eqref{mixedrindler} obeys
\be\label{mixedrindler2}
\lan \phi_L\wt{\phi_R}|\Omega\ran\propto \lan \wt{\phi_R}|e^{-\pi \hat{K}_R}i^{F_R}\hat{\Theta}_\CRT^L|\phi_L\ran
\ee
with
\be
\hat{K}_{R}=\int_0^\infty d\hat{x} \hat{x} T_{\hat{t}\hat{t}}
\ee
the right boost operator and $\hat{\Theta}_{\CRT}^L$ the restricted $\CRT$ operator for the plane (we are using hats to distinguish coordinates and operators on the plane from coordinates and operators on the cylinder). 
Indeed taking $\hat{t}=-i \hat{t_E }$ and defining $w=\hat{x}+i\hat{t_E }$, a conformal transformation that maps the interval $(0,\ell)$ on a circle of circumference $R$  to the region $x>0$ and the interval $(\ell-R,0)$ to $x<0$ is
\be
w(z)=\frac{R\sin \left(\frac{\pi}{R}r\right)\sin\left(\frac{\pi}{R}z\right)}{\pi\sin\left(\frac{\pi}{R}(\ell-z)\right)}.
\ee
Here $z=x+it_E $ are complex coordinates on the circle with circumference $R$.
Under this conformal transformation the modular operator $K_R$ in the $w$ coordinates maps to\footnote{Here we suppress an additive constant in the modular operator $K$ which arises from the conformal anomaly, this rescales $\hat{\rho}$ by a constant but we anyways are not trying to keep track of its normalization.}
\begin{align}\nonumber
K&=\int_0^\ell dx \frac{w(x)}{w'(x)}T_{tt}(x)\\
&=\int_0^\ell dx \frac{\sin \left(\frac{\pi}{R}x\right)\sin\left(\frac{\pi}{R}(\ell-x)\right)}{\frac{\pi}{R}\sin\left(\frac{\pi}{R}\ell\right)}T_{tt}(x)
\end{align}
and the $\CRT$ operator $\hat{\Theta}_{\CRT}$ maps to an operator $\Theta_{\CRT}$ which (in Euclidean signature) implements the diffeomorphism
\be
e^{2\pi i z'/R}=\frac{e^{2\pi i z/R}+e^{2\pi i(z-\ell)/R}-2}{2e^{2\pi i (z-\ell)/R}-e^{-2\pi i\ell/R}-1}
\ee
and in Lorentzian signature takes the complex conjugate of all fields as well.  In particular we are interested in the case where $R=2L$ and $\ell=L$, in which case these expressions simplify to
\begin{align}\nonumber
w(z)&=\frac{2L}{\pi}\tan\left(\frac{\pi z}{2L}\right)\\\nonumber
K&=\int_{0}^L dx\frac{L}{\pi}\sin\left(\frac{\pi}{L}x\right)T_{tt}(x)\\
z'&=-z.
\end{align}
The image in the $z$ plane of the modular $\pi$ rotation used in constructing the Rindler decomposition of the vacuum on the circle is shown in figure \ref{crtcylfig}.
\bfig
\includegraphics[height=7cm]{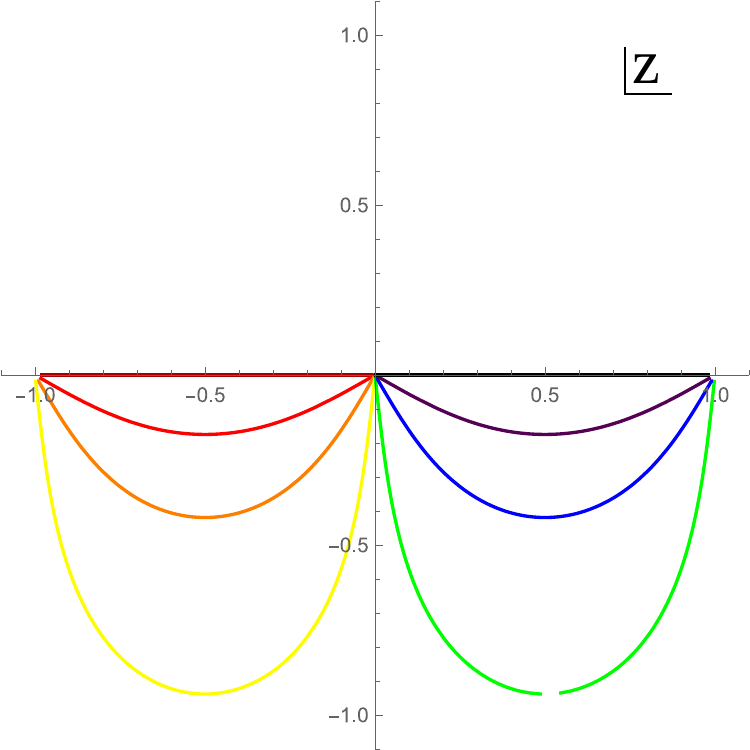}
\caption{The rotation by $\pi$ in Euclidean modular time that prepares the ground state on a circle.  The rotation takes the dark red slice to the dark purple slice passing through the color wheel, and the full rotation is the natural definition of $\CRT$ on the circle.  Here we have set $L=1$ so the circle has circumference two.}\label{crtcylfig}
\efig

Using this conformal transformation and equation \eqref{mixedrindler2}, from figure \ref{mobcylfig} we have
\be
\lan \wt{\phi}_f|\hat{\rho}|\phi_i\ran\propto\lan \wt{\phi}_f|e^{-\pi K_R}i^{F_R}\Theta_{\CRT}^LU_{trans}|D^*_\T\phi_i^*\ran,
\ee
where we have included a unitary $U_{trans}$ that implements a translation by $L$ to move $\phi_i$ from the ``right'' part of the cylinder with $0<x<L$ to the ``left'' part with $-L<x<0$ as in equation \eqref{t0conds}.  Finally we observe that
\be
|D_\T^*\phi_i^*\ran=i^F\Theta_\T|\phi_i\ran,
\ee
as we can check by acting on the right-hand side with $\Phi$:
\begin{align}\nonumber
\Phi(x)i^F\Theta_\T|\phi_i\ran&=i^F \Theta_\T D_\T\Phi(\T x)|\phi_i\ran\\
&=D_\T^*\phi_i^*(\T x)i^F\Theta_\T|\phi_i\ran,
\end{align}
and therefore we have
\be
\lan \wt{\phi}_f|\hat{\rho}|\phi_i\ran\propto\lan \wt{\phi}_f|e^{-\pi K_R}i^{F_R}\Theta_{\CRT}^LU_{trans}i^{F_R}\Theta_\T|\phi_i\ran
\ee
and thus
\be\label{Mobens}
\hat{\rho}=e^{-\pi K_R}(-1)^{F_R}\Theta_{\CRT}^L U_{trans} \Theta_\T.
\ee
Here we have used that $i^{F_R}$ commutes with $\Theta_{\CRT}^LU_{trans}$.  The unitary operator $\Theta_{\CRT}^L U_{trans} \Theta_\T$ implements a cylinder version of $\C\R$.  We emphasize that $\hat{\rho}$ is linear (as opposed to antilinear), but it is not positive or even hermitian; instead it is a postive operator times a unitary operator.

%
%

\section{Gauging inversions and the sum over topology}\label{sumsec}
Having introduced background gauge fields for spacetime inversions, the natural next step is to make these gauge fields dynamical by summing over them in the path integral.  In quantum field theory this is not such a natural thing to do: the subgroups of the Lorentz group $O(d-1,1)$ which are generated by $\R$ and/or $\T$ are not normal subgroups, so gauging one of these subgroups necessarily breaks Lorentz invariance (this was recently emphasized in \cite{McNamara:2022lrw}).  It is therefore more appealing to gauge $\R$ and/or $\T$ in theories where the full identity component $SO^+(d-1,1)$ of the Lorentz group is also gauged, namely in theories with dynamical gravity.  To make our presentation manageable we will discuss things in terms of $\R$ and $\C\T$ instead of $\R$ and $\T$, since as discussed at the end of section \ref{topsec} the former are related by a rotation in Euclidean signature and so fit together more naturally.  

In this section we will consider gravitational theories where all dynamical fields are tensors, in which case the natural mathematical language for discussing the gauging of $\R$ and/or $\C\T$ is the structure group of the tangent bundle.   A general differentiable manifold $M$ of dimension $d$ is covered by patches $U_\alpha$, with transition functions 
\be
x_\alpha^\mu=f_{\alpha\beta}^\mu(x_\beta)
\ee
such that the $d\times d$ matrices
\be
(D_{\alpha\beta})^\mu_{\phantom{\mu}\nu}=\frac{\partial f_{\alpha\beta}^\mu}{\partial x_\beta^\nu},
\ee
which are used to define the transformations of tensor fields from one patch to another, are invertible.    We therefore say that the tangent bundle has structure group $GL(d,\mathbb{R})$.  When $M$ is equipped with additional structure such as a metric, we can try to find special sets of patches where the $D_{\alpha\beta}$ obeys additional restrictions beyond invertibility; this procedure is called \textit{reduction of the structure group}.  The gauging of $\R$ and/or $\C\T$ in a gravitational theory is a statement about far we can reduce the structure group of the tangent bundle for the allowed set of spacetimes in the path integral.   

We note in passing that the matrices $D_{\alpha\beta}$ are not sufficient to define the transformations of spinor fields from one patch to another, so defining fermionic theories on manifolds requires additional structure.  We'll discuss the necessary generalization in section \ref{fermsec}.  Another generalization is to include dynamical gauge fields for internal symmetries, in which the fields become sections of vector bundles or connections on principal bundles, but as this is more well-known and also less relevant we will not discuss it. 

\bfig
\includegraphics[height=6cm]{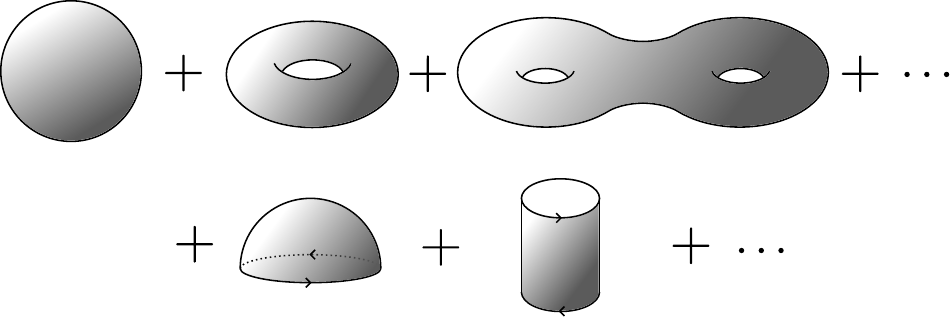}\caption{Organizing the sum over two-dimensional closed Euclidean geometries by structure group: with structure group $SO(2)$ we include only the top row, while with structure group $O(2)$ we also include the bottom row.  In Lorentzian signature these correspond to gauging $\R\T SO(1,1)$ and $O(1,1)$ respectively, so the bottom row should be included only if versions of both $\R$ and $\C\T$ are gauge symmetries.}\label{genusfig}
\efig
Let's first see how the reduction of the structure group works in Euclidean signature, where $M$ comes equipped with a Riemannian metric $g_{\mu\nu}$.  One consequence of having such a metric is that in each sufficiently small coordinate patch $U_\alpha$ we can choose an orthonormal frame $\{e_a\}$, meaning a set of $d$ orthonormal vectors $e_a^{\phantom{a}\mu}(x)$ that varies continuously throughout $U_\alpha$.  Here the symbol $e_a^{\phantom{a}\mu}(x)$ means the $\mu$th component of the $a$th vector in the basis (we have suppressed the dependence on the patch $\alpha$, but it is important to remember that the frame field also depends on this).  $e_a^{\phantom{a}\mu}(x)$ is conventionally referred to as the vielbein.    Orthonormality implies that we have
\be
e_a^{\phantom{a}\mu}e_b^{\phantom{a}\nu}g_{\mu\nu}=\delta_{ab}
\ee
and also
\be
e^a_{\phantom{a}\mu}e^b_{\phantom{b}\nu}\delta_{ab}=g_{\mu\nu},
\ee
where we have defined the inverse vielbein
\be
e^a_{\phantom{a}\mu}(x)=\delta^{ab}g_{\mu\nu}(x)e_b^{\phantom{a}\nu}(x).
\ee
We can turn any vector with a coordinate index $\mu$ into a vector with a frame index $a$ by contracting with the inverse vielbein
\be
V^a(x)=e^a_{\phantom{a}\mu}V^\mu,
\ee
and we can just as well write the transformation of $V$ in terms of frame indices:
\be
(V_\alpha)^a=(D_{\alpha\beta})^a_{\phantom{a}b}(V_\beta)^b.
\ee
Orthonormality of the frame $\{e_a\}$ implies that the transformation matrix $D_{\alpha\beta}$ obeys (suppressing frame indices)
\be
(D_{\alpha\beta})^TD_{\alpha\beta}=I,
\ee
or in other words that $D_{\alpha\beta}$ (with frame indices) is an element of the Euclidean rotation group $O(d)$.  We thus have reduced the structure group from $GL(d,\mathbb{R})$ to $O(d)$.  For a general Riemannian manifold no further reduction is possible.  If $M$ is orientable however, then by choosing an orientation on $M$ we can further restrict our choice of vielbein $e_a^{\phantom{a}\mu}$ to be compatible with this orientation. The matrices $D_{\alpha\beta}$ will then be elements of $SO(d)$, which further reduces the structure group to $SO(d)$.  The converse of this statement is also true: a Riemannian manifold $M$ is orientable if and only if we can reduce the structure group to $SO(d)$. The question of whether spatial reflection is a gauge or global symmetry in Euclidean signature is precisely the question of whether or not we allow unoriented Riemannian manifolds in the Euclidean path integral (see figure \ref{genusfig} for an illustration).  This is nicely compatible with the nontrivial $\R$ backgrounds we constructed in section \ref{Rsubsec} above.\footnote{In section \ref{Rsubsec} we also saw that gauging $\R$ can lead to $\R$-symmetric boundaries in spacetime.  On the string world sheet these are the endpoints of open strings as we discussed above, while for reflections in the target space as we consider here they might more properly be called ``end-of-the-world branes''.  The existence of such branes when parity of is gauged was recently proposed in \cite{McNamara:2022lrw}, and we discuss it further in section \ref{comsec} below.}    

Turning to Lorentzian signature, the discussion of the last paragraph proceeds in the same way except that the frame fields now obey
\begin{align}\nonumber
e_a^{\phantom{a}\mu}e_b^{\phantom{a}\nu}g_{\mu\nu}&=\eta_{ab}\\
e^a_{\phantom{a}\mu}e^b_{\phantom{b}\nu}\eta_{ab}&=g_{\mu\nu},
\end{align}
where $\eta_{ab}$ is the Minkowski metric, and the constrant on $D_{\alpha\beta}$ becomes (again suppressing frame indices)
\be
(D_{\alpha\beta})^T \eta D_{\alpha\beta}=\eta.
\ee
Thus a Lorentzian metric allows us to reduce the structure group of $M$ from $GL(d,\mathbb{R})$ to the Lorentz group $O(d-1,1)$.  As we discussed in section \ref{topsec}, $O(d-1,1)$ has four connected components and it is natural to ask whether or not we can further reduce the structure group to some subset of these.  To discuss this it is convenient to give names and interpretations to the various options:
\bi
\item $SO^+(d-1,1)$: the identity component of $O(d-1,1)$.  Sometimes called the proper orthochronous Lorentz group.  Spacetimes with this structure group are both orientable and time-orientable, so restricting to this structure group means that neither $\R$ nor $\C\T$ is gauged.  
\item $\R SO(d-1,1)$: the subgroup generated by the identity component and $\R$.  Sometimes called the orthochronous Lorentz group. Spacetimes with this structure group are time-orientable but not necessarily orientable, so $\R$ is gauged but $\C\T$ is not.  
\item $\T SO(d-1,1)$: the subgroup generated by the identity component and $\T$.  Spacetimes with this structure group need neither be orientable nor time-orientable, and $\C\T$ is gauged but $\R$ is not.
\item $SO(d-1,1)$: the subgroup generated by the identity component and $\R\T$.  Sometimes called the proper Lorentz group.  Spacetimes with this structure group are orientable but not necessarily time-orientable. $\C\R\T$ is gauged but neither $\R$ nor $\C\T$ is gauged.
\item $O(d-1,1)$: the full Lorentz group, generated by the identity component, $\R$, and $\T$.  Spacetimes with this structure group need neither be orientable nor time-orientable, and $\R$ both and $\C\T$ are gauged.
\ei

It is perhaps curious that in Euclidean signature we had only two choices for the reduced structure group, $SO(d)$ and $O(d)$, while in Lorentzian signature we have five.  The origin of this discrepancy is that the topological fact we reviewed at the beginning of section \ref{topsec}: the Euclidean continuation of $\C\R\T$ is a rotation by $\pi$, and this is in the identity component of $O(d)$.  Since in gravity we always gauge this identity component, we see that the only Lorentzian options which arise naturally by analytic continuation from Euclidean signature are those for which $\C\R\T$ is a gauge symmetry.  These are the last two: $SO(d-1,1)$ and $O(d-1,1)$ (see figure \ref{genusfig}).  If $\R$ and $\C\T$ are not separately symmetries, as seems to be the case in our world (with the possible exception of them being gauged and then spontaneously broken), than $SO(d-1,1)$ is the only option.  Either way, time non-orientable geometries much be included.

It is instructive to consider what happens in Euclidean signature if we try to analytically continue one of the other three choices for the Lorentzian gauge group.  For example from a Lorentzian path integral point of view nothing stops us from treating $\CRT$ as a global symmetry.   Since this continues to a rotation by $\pi$ in Euclidean signature, and we can obtain such a rotation by combining rotations by $\pi/n$ for any integer $n$ about any axis, this gives a severe restriction on the possible transition functions, essentially requiring that they are all trivial.  As an illustration of the severity of such a constraint, in appendix \ref{windingapp} we show that for any  closed oriented two-dimensional surface $\Sigma$ the Euler characteristic obeys the following formula:
\be\label{winding}
\chi(\Sigma)=\sum_{\alpha<\beta<\gamma}\int_{V_{\alpha\beta\gamma}}w_{\alpha\beta\gamma},
\ee
where $V_{\alpha\beta\gamma}$ are the triple intersections of any triangulation $V_\alpha$ of $\Sigma$ and $w_{\alpha\beta\gamma}$ counts how many times the conformal transition functions wind around $V_{\alpha\beta\gamma}$ as we change frames $V_\alpha\to V_\beta\to V_\gamma \to V_\alpha$.  This shows that as long as the Euler characteristic is not zero, it is necessary for the transition functions from one patch to another to include nontrivial rotations.  Any rule which required the vanishing of such rotations would introduce a strong global restriction on surfaces: they would need to have vanishing Euler characteristic.  Such a requirement does not seem compatible with locality.  


\section{An example from AdS/CFT}\label{adssec}
So far our discussion has been somewhat formal, so we now bring it down to earth by discussing a concrete example of a nontrivial $\CRT$ gauge field in the context of the $AdS_3/CFT_2$ correspondence.  The example we will consider is a quotient of the two-sided BTZ geometry.  In Kruskal coordinates the BTZ geometry \cite{Banados:1992wn} has the metric
\be\label{BTZ}
ds^2=-\frac{4\, dX^+dX^-}{(1+X^+X^-)^2}+\left(\frac{2\pi}{\beta}\cdot\frac{1-X^+X^-}{1+X^+ X^-}\right)^2d\phi^2,
\ee
where $\beta$ is the inverse temperature and $\phi\in (0,2\pi)$ is the angular coordinate.  We are working in units where the AdS radius is one.  Continuing to Euclidean signature we have
\be\label{eBTZ}
ds^2=\frac{4\, dz d\ol{z}}{(1-|z|^2)^2}+\left(\frac{2\pi}{\beta}\cdot\frac{1+|z|^2}{1-|z|^2}\right)^2d\phi^2,
\ee
with $0\leq|z|\leq 1$.  The thermal partition function is obtained by evaluating the Euclidean Einstein-Hilbert action
\be
S=-\frac{1}{16\pi G}\int \sqrt{g}d^3x (R+2)-\frac{1}{8\pi G}\int \sqrt{\gamma}d^2x K
\ee
on this solution, which after a holographic renormalization gives
\be\label{ZBTZ}
Z(\beta)\approx \exp\left(\frac{\pi^2}{2G\beta}\right)=\exp\left(\frac{\pi^2 c}{3\beta}\right).
\ee
Here 
\be
c=\frac{3}{2G}
\ee
is the Brown-Henneaux central charge of $AdS_3$ \cite{Brown:1986nw}.  This formula agrees with the Cardy formula \cite{Cardy:1986ie} for the high-temperature thermal partition function of an arbitrary two-dimensional CFT with central charge $c$ \cite{Strominger:1997eq}.  

\bfig
\includegraphics[height=6cm]{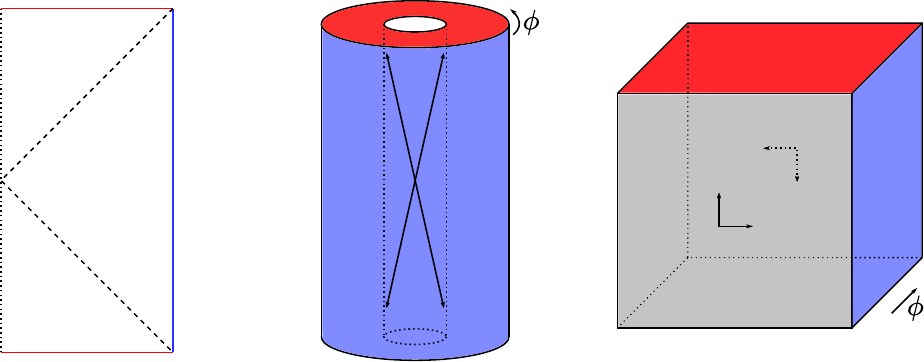}
\caption{Three representations of the $\CRT$-twisted black hole.  In each case the $AdS_3$ boundary is shaded blue and the singularity is shaded red (despite appearances the ``future'' and ``past'' singularities are connected).  On the left is a Penrose diagram-type representation, which covers half of the BTZ Penrose diagram.  The dotted line in the center represents a cylinder with an antipodal identification, i.e. a copy of the Lorentzian M\"obius strip.  In the center we give a three-dimensional representation of the same idea: the spacetime lies between the two vertical cylinders, with the outer cylinder representing the (single) $AdS_3$ boundary and the interior cylinder (representing $X=0$) being antipodally identified to produce a smooth spacetime.  On the right we give an alternative representation where we choose a different fundamental domain: the front and back faces of the cube are identified with a rotation by $\pi$.}\label{CRTtwistfig}
\efig
The geometry we are interested in is obtained from the BTZ geometry by the identification
\be
(X^+,X^-,\phi)\sim (-X^+,-X^-,\phi+\pi),
\ee
or equivalently  
\be
(T,X,\phi)\sim (-T,-X,\phi+\pi)
\ee
if we define
\be
X^\pm=T\pm X.
\ee  This identification has no fixed points, so the resulting geometry is a smooth Lorentzian manifold.  We will refer to it as a ``$\CRT$-twisted black hole''.\footnote{This geometry differs from the well-known ``$\mathbb{RP}_2$ geon'' geometry (see e.g. \cite{Misner:1957mt,Louko:1998hc,Maldacena:2001kr}), which would have instead identified
\be
(X^+,X^-,\phi)\sim (X^-,X^+,\phi+\pi).
\ee
The four-dimensional Schwarzschild version of the $\CRT$-twisted black hole was previously considered by Gibbons in \cite{Gibbons:1986rzu}, who argued that it requires modifications of quantum mechanics on grounds which boil down to the lack of a natural Euclidean state.  As in our discussion of the Lorentzian M\"obius strip however, the lack of a natural Euclidean state does not imply a modification of quantum mechanics.  And indeed we will see shortly that the $\CRT$-twisted black hole has a fairly conventional interpretation in the dual CFT.
}
The $\CRT$-twisted black hole has a single asymptotically-$AdS_3$ boundary, whose topology is the usual Lorentzian cylinder $\mathbb{S}^1\times \mathbb{R}$.  We therefore should be able to assign it some kind of interpretation in the standard Hilbert space of the dual CFT on a spatial circle.  On the other hand it is \textit{not} time-orientable.  One way to see this is to note that the $X=0$ slice has the topology of the Lorentzian M\"obius strip, so there are spatial circles with nontrivial $\T$-holonomy.  The global topology is $\mathbb{S}^1\times \mathbb{R}^2$, with a $\CRT$ transformation of the $\mathbb{R}^2$ being implemented as we go around the circle.  We give a few graphical representations of the $\CRT$-twisted black hole in figure \ref{CRTtwistfig}.

\bfig
\includegraphics[height=6cm]{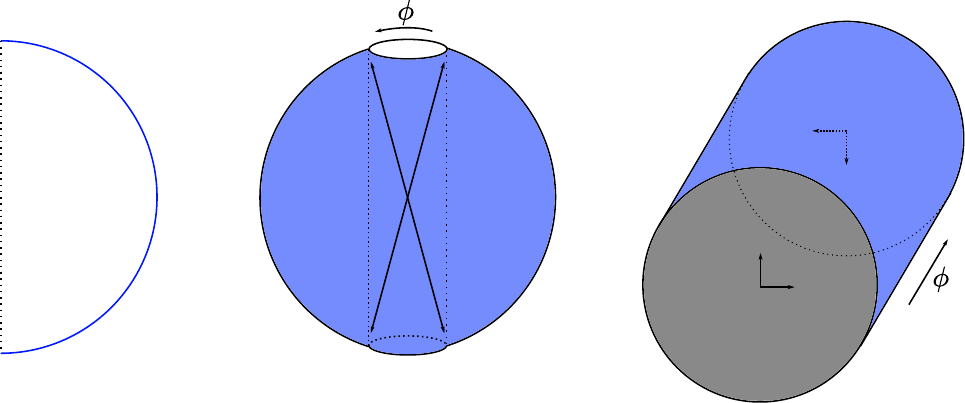}
\caption{Three representations of the Euclidean $\CRT$-twisted black hole.}\label{eCRTtwistfig}
\efig
As in our discussion of quantum field theory on the Lorentzian M\"obius strip, there is no natural state for perturbative gravity around the $\CRT$-twisted black hole.  We can again instead use the Euclidean path integral to prepare a natural but non-positive pseudostate $\hat{\rho}$, and then study correlation functions in this pseudostate.  The pseudostate to use here is prepared by the Euclidean continuation of $\CRT$-twisted black hole, which is obtained from the Euclidean BTZ geometry \eqref{eBTZ} by the identification
\be
(z,\ol{z},\phi)\sim(-z,-\ol{z},\phi+\pi).
\ee
This geometry is shown in figure \ref{eCRTtwistfig}; its topology is that of the solid torus $\mathbb{S}^1\times B^2$, with a rotation by $\pi$ of the $B^2$ as we go around the $\mathbb{S}^1$.  The $AdS_3$ boundary therefore has the topology of the torus $\mathbb{S}^1\times \mathbb{S}^1$.  Inspecting the left diagram of figure \ref{eCRTtwistfig}, we see that this torus has a spatial rotation by $\pi$ as we go around a thermal circle of radius $\frac{\beta}{2}$.  The CFT Hilbert space interpretation of the Euclidean partition function on this geometry is therefore
\be\label{ZCFT}
Z_{CFT}=\Tr\left(e^{-\frac{\beta}{2}H}e^{i\pi J}\right),
\ee
where $J$ is the angular momentum on the spatial circle of the CFT.  We thus have learned that the CFT dual of the $\CRT$-twisted black hole pseudostate is
\be\label{CFTens}
\hat{\rho}=e^{-\frac{\beta}{2}H}e^{i\pi J},
\ee
which is quite similar to our expression \eqref{Mobens} for the M\"obius pseudostate on the Lorentzian M\"obius strip in field theory.  

For this example it is actually possible to give a direct match between the gravitational action of the Euclidean $\CRT$-twisted black hole and the CFT trace of the pseudostate \eqref{CFTens}.  On the bulk side the geometry is a $\mathbb{Z}_2$ quotient of the Euclidean BTZ geometry, so its Euclidean action is just half that of the BTZ black hole (with inverse temperature $\beta$). From \eqref{ZBTZ} we therefore have
\be\label{Zbulk}
Z_{gravity}=\exp\left(\frac{\pi^2 c}{6\beta}\right).
\ee
To match this in the dual CFT we can use modular invariance.  The twisted trace \eqref{ZCFT} is equivalent (see equation \eqref{taudef}) to a torus partition function with modular parameter
\be
\tau=\frac{1}{2}+i\frac{\beta}{4\pi}.
\ee
We are interested in studying this in the limit of small $\beta$, so defining $\epsilon=\frac{\beta}{4\pi}$ we can use a sequence of modular transformations
\begin{align}
Z_{CFT}(\frac{1}{2}+i\epsilon,\frac{1}{2}-i\epsilon)&=Z_{CFT}\left(-\frac{1}{\frac{1}{2}+i\epsilon},-\frac{1}{\frac{1}{2}-i\epsilon}\right)\\\nonumber
&\approx Z_{CFT}(-2+4 i\epsilon,-2-4 i \epsilon)\\\nonumber
&=Z_{CFT}(4 i\epsilon,-4 i \epsilon)\\\nonumber
&=Z_{CFT}\left(\frac{i}{4  \epsilon},-\frac{i}{4\epsilon}\right)
\end{align}
to put this partition function into a low-temperature limit where it is dominated by the Casimir energy (this is a generalized version of the derivation of the Cardy formula).  Indeed the partition function with $\tau=\frac{i}{4\epsilon}$ describes a transformed torus with inverse temperature $\hat{\beta}$ and spatial circumference $\hat{L}$ obeying
\be
\frac{\hat{\beta}}{\hat{L}}=\frac{1}{4\epsilon}=\frac{\pi}{\beta}.
\ee
The Casimir energy for any (nonchiral) 2D CFT with central charge $c$ on a spatial circle of circumference $\hat{L}$ is given by\footnote{This result is derived by computing the transformation of the stress tensor under the conformal transformation from the plane to the cylinder.}
\be
\hat{E}_{casimir}=-\frac{\pi c}{6\hat{L}},
\ee
so we thus have
\begin{align}\nonumber
Z_{CFT}&\approx \exp\left(-\hat{\beta}\hat{E}_{casimir}\right)\\\nonumber
&=\exp\left(\frac{\pi \hat{\beta}c}{6\hat{L}}\right)\\
&=\exp\left(\frac{\pi^2 c}{6\beta}\right)
\end{align}
in perfect agreement with the gravitational result \eqref{Zbulk}.\footnote{Some readers may wonder why we did not interpret the partition function \eqref{ZCFT} as arising from a rotating BTZ geometry with an imaginary chemical potential for angular momentum.  This is indeed another valid saddle point for these boundary conditions, but it is subleading to the $\CRT$-twisted black hole.}  This calculation shows that the Euclidean $\CRT$-twisted black hole is the dominant contribution to the path integral, and we view it as providing direct evidence that $\CRT$ is indeed a gauge symmetry on the gravitational side of the AdS/CFT correspondence.  

\bfig
\includegraphics[height=7cm]{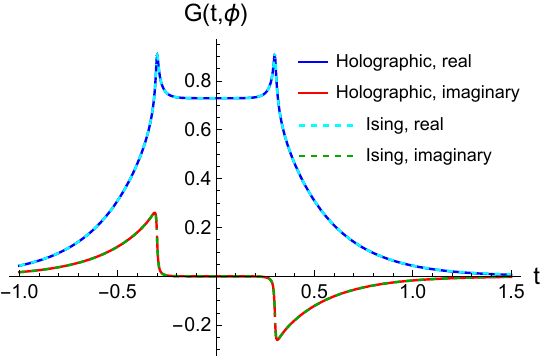}
\caption{Plots of Wightman functions in  \eqref{HWightman} and \eqref{IsingWightman}.
The parameters are set to  $\beta = 0.2$ , $\epsilon = 0.005$ and $\phi_2 - \phi_1 = 0.3$.}\label{TTcorrelatorsfig}
\efig
We can also confirm the necessity of including the $\CRT$-twisted black hole directly in Lorentzian signature by comparing Wightman functions.  Correlation functions of free fields in the $\CRT$-twisted black hole pseudostate with inverse temperature $\beta$ can be computed using the method of images as in our discussion of the Lorentzian M\"obius strip.  In particular the boundary Wightman correlator of a scalar operator $O$ of dimension $\Delta$ on a spatial circle of radius one is given by
\begin{align}\nonumber
\lan O(t_2,\phi_2) O(t_1,\phi_1)\ran=&\sum_{n=-\infty}^\infty\Bigg[\frac{1}{\left(-\cosh\left(\frac{2\pi}{\beta}(t_2-t_1-i\epsilon)\right)+\cosh\left(\frac{2\pi}{\beta}(\phi_2-\phi_1+2\pi n)\right)\right)^{\Delta}}\\
&+\frac{1}{\left(\cosh\left(\frac{2\pi}{\beta}(t_2-t_1)\right)+\cosh\left(\frac{2\pi}{\beta}\left(\phi_2-\phi_1+2\pi(n+1/2)\right)\right)\right)^{\Delta}}\Bigg], \label{HWightman}
\end{align}
with the first term being the BTZ result \cite{Keski-Vakkuri:1998gmz,Maldacena:2001kr} and the second term coming from an image charge on the other boundary with opposite time and shifted by $\pi$ in the $\phi$ direction.  Note that the second term does not give rise to any singularities, so this correlator is consistent with boundary causality. One could try to match it to an analytic boundary calculation based on assuming the dominance of the thermal Virasoro identity block, but we instead will simply compare it numerically to the (presumably universal) high-temperature torus correlator of the Ising CFT (meaning the free Majorana fermion with gauged fermion parity that we considered at the end of section \ref{intsec}) with $\tau=\frac{1}{2}+\frac{\beta}{4\pi}$.
The Ising spin-spin correlator on the torus in Euclidean signature is given by \cite{DiFrancesco:1997nk}:
\begin{align}
\lan \sigma(z_2,\bar{z}_2)\sigma(z_1,\bar{z}_1) \ran = \Big| \frac{\partial_{z_{2}} \Theta_1(z_2 - z_1 | \tau)|_{z_2=z_1}}{\Theta_1 (z_2 - z_1 | \tau)} \Big| ^{\frac{1}{4}} \frac{\sum_{\nu = 1}^4 |\Theta_{\nu}(\frac{z_2-z_1}{2}|\tau)|}{\sum_{\nu=2}^4 |\Theta_{\nu} (0|\tau)|}, \label{IsingWightman}
\end{align}
where $z_i = \phi_i + i \tau_i$ denote coordinates on a torus and the functions $\Theta_\nu(z|\tau)$ are defined in section 10.A.2 of \cite{DiFrancesco:1997nk} (they are two-parameter generalizations of the functions defined by \eqref{thetadef}).
To obtain a Wightman function in the pseudostate \eqref{CFTens}, we perform the analytic continuation $\tau_1 = - i t_1$ and $\tau_2 = -it_2 + \epsilon  $.
In figure \ref{TTcorrelatorsfig} we compare the Wightman function \eqref{HWightman} obtained from the $\CRT$-twisted black hole with the corresponding Wightman function \eqref{IsingWightman} of the Ising CFT.\footnote{In \eqref{HWightman} we set $\Delta=\frac{1}{8}$ to match the scaling dimension of the spin operator and also  multiply by $2^{\Delta} (\frac{\pi}{\beta})^{2\Delta}$ to match the normalization of \eqref{IsingWightman} in the OPE limit $z_2 \to z_1$.} At high temperatures, where we expect universal behavior, we see the two are in good agreement.

It is interesting to consider how this example generalizes to higher dimensions.  The idea would be to take the AdS-Schwarzschild geometry and perform the identification 
\be
(T,X,\Omega)\sim(-T,-X,-\Omega),
\ee
where $\Omega \to -\Omega$ indicates the antipodal map on $\mathbb{S}^{d-2}$.  This again produces a smooth one-sided geometry, with a natural Euclidean pseudostate in the dual CFT given by
\be\label{dpseudo}
\hat{\rho}=e^{-\beta H/2}U_{AP}.
\ee
Here $U_{AP}$ is the unitary operator which implements the antipodal map in the dual CFT.  When $d$ is even this map is orientation-reversing, so this geometry and pseudostate only make sense in theories where $\R$ is a gauge symmetry.  In principle one could attempt to confirm this explicitly in the dual CFT as we have done for $d=3$ in this section, but Cardyology is less powerful in higher dimensions so other techniques would be needed.  This antipodal gluing has been studied by 't Hooft, see for example \cite{tHooft:2016qoo}.  We think that the pseudostate \eqref{dpseudo} gives a correct interpretation of this geometry.  The gauging of CRT in this context has been discussed from a somewhat different point of view in \cite{Betzios:2020wcv,Betzios:2020xuj}.

\section{Chronological censorship}\label{censorsec}
The inclusion of geometries that are not time-orientable on the bulk side of the AdS/CFT correspondence may be concerning.  Usually time-orientability is seen as necessary for preserving causality (see e.g. chapter eight of \cite{Wald:1984rg}), and in AdS/CFT we know that boundary causality must be exactly preserved.  Therefore any causal pathologies in the gravitational description arising from lack of time-orientability must be strictly invisible to boundary observers.  For the $\CRT$-twisted black hole this is indeed the case: each point on the dotted cylinder in figure \ref{CRTtwistfig} lies on or behind the past or future horizon, and so cannot be reached by a causal curve that starts and ends at the AdS boundary.  Such curves therefore have the same causal properties as in the BTZ geometry, and in particular are compatible with boundary causality (as we just saw for the boundary two-point function).  Another way to think about this is that although this geometry has self-intersecting timelike curves (SITCs), their self-intersections always lie behind the past or future horizon.  This is fortunate, since given such an intersection outside of both horizons we could use it to violate boundary causality by sending in a causal curve from the boundary and either having it enter and exit a closed timelike curve (if the tangent vectors at the self-intersection have the same local time orientation) or having it come back out going backward in time (if the tangent vectors at the self-intersection have opposite local time orientation).  This suggests a general notion that we will call \textit{chronological censorship}: SITCs are allowed in quantum gravity, but only when their self-intersections lie behind horizons.  This specific idea so far does not seem to have been proposed in the relativity literature, but for a review of related issues see \cite{Friedman:2006tea}. One possible chain of logic arguing for chronological censorship of SITCs arising from the gauging of $\CRT$ (or $\T$ or $\C\T$) is the following:  if we wish to avoid naked singularities and obvious problems with boundary causality, then we probably want to assume that the region exterior to any horizons, usually called the domain of outer communication (DOOC), is globally hyperbolic (this is sometimes called strong asymptotic predictability).  The topological censorship theorem \cite{Friedman:1993ty,galloway1995topology} then says that the DOOC must be simply connected, and therefore that any spatial loops with nontrivial $\CRT$ (or $\T$/$\C\T$) holonomy must be at least partially hidden behind the horizons.\footnote{There seems to be some disagreement in the relativity literature about exactly what assumptions are necessary for topological censorship to hold, see e.g. \cite{krasnikov2013topological,Chrusciel:2019lqq}, which we will not try to disentangle.}  In order to have an SITC we need to have a timelike curve that completes a journey around one of these loops, but that means that the piece of this curve which is ``between'' a self-intersection cannot lie entirely in the DOOC.  Since part of this piece is behind the past and/or the future horizon, it means that its entire past and/or its entire future along the curve, and thus in particular the self-intersection point since is in both of these, must lie behind the past and/or the future horizon.\footnote{As a first check of chronological censorship we can consider perturbing the $\CRT$-twisted black hole by adding a null spherical shell of matter (of positive energy) which bounces off of the AdS boundary at $t=0$. The backreacted geometry is obtained by gluing pieces of the BTZ geometry with slightly higher mass to the exterior of the shell, which has the effect of moving the identified cylinder further into the interior relative to the new horizons.  This takes us from a situation where boundary causality can almost be violated to one where it is protected with a finite margin of error.}

\section{Quantum mechanics in a closed universe}\label{quantsec}
The existence of a time-reversing gauge symmetry has a profound consequence for the quantum gravity of a closed universe such as de Sitter space: in a closed universe all physical states must be gauge-invariant, and the subspace of any Hilbert space $\mathcal{H}$ which is invariant under an antiunitary operator $\Theta$ is a \textit{real} vector space.\footnote{Here is the proof. Let $|\psi\ran,|\phi\ran\in \mathcal{H}$ be linearly-independent pure states which are each invariant under $\Theta$.  Acting on a general superposition we have
\be
\Theta\left(a|\psi\ran+b|\phi\ran\right)=a^*|\psi\ran+b^*|\phi\ran,
\ee
which by linear independence is equal to $a|\psi\ran+b|\phi\ran$ if and only if $a$ and $b$ are real.  We give a more explicit description of the set of states invariant under an antiunitary operator $\Theta$ at the end of appendix \ref{alapp}.}  Since $\CRT$ is a time-reversing gauge symmetry that always exists, we thus learn that the Hilbert space of quantum gravity in a closed universe is real.  This conclusion may seem implausible.  After all many of the most distinctive consequences of quantum mechanics, for example the interference pattern in the double slit experiment or the ability of quantum computers to factor integers in polynomial time using Shor's algorithm, seem to rely crucially on having superpositions of states with complex coefficients.  If the physical Hilbert space of the universe is real, does this mean that we need to reject all of these phenomena?  Or does it (more plausibly) mean that we need to reject one of the assumptions that led to this conclusion?  Of course the answer is no to both of these questions, and we now explain how conventional quantum mechanics can arise in a real vector space.    

The key idea we will need is that measuring time in a closed universe requires some kind of clock.  This may sound like a triviality, and indeed on some level it is, but in most discussions of quantum mechanics the issue is suppressed by always thinking from the point of view of an observer who is external to the system being considered.  The need for a physical clock in cosmological applications of quantum gravity has been understood for a long time, see for example \cite{Page:1983uc,Marolf:1994nz,Maldacena:2002vr,Witten:2023xze} for discussions from various points of view.  Here we will motivate the idea in a mundane way by analogy to electrodynamics.  Indeed say that we lived in a closed universe without gravity, which for concreteness we can take to be a round spatial $\mathbb{S}^3$ of radius $L$.  Moreover let's say that $L$ is very large compared to anything of direct relevance for us, such as people, planets, stars, etc.  A natural question to consider in this situation would be the following: if we collide two electrons together, what comes out?  After all we have done this experiment many times in our own universe, and we have learned a lot from it.  But in the $\mathbb{S}^3$ universe there is an apparent problem: we are unable to prepare an initial scattering state with two electrons in an otherwise empty universe, since in a closed universe the total electric charge must be zero (this follows by integrating Gauss's law over all of space).  The solution to this problem is very well-known (see e.g. \cite{Dirac:1955uv}): in order to have two electrons here, there also needs to be something with electric charge two, say two positrons, somewhere else in the universe.  The electric flux lines from our electrons end on these positrons, giving a consistent solution of Gauss's law.  In the limit of large $L$ the positrons can be very far away from the electrons, and thus have only a very small effect on the result of the collision.  In this example the positrons are playing the same role as the clock does in quantum cosmology. 

To make this concrete, let's first consider the situation where our time-reversing antiunitary symmetry $\Theta$ obeys $\Theta^2=1$ (as is the case for $\CRT$).  We'll assume that our full system is the tensor product of a two-state clock system $C$ and some additional system $S$ on which we would like to do measurements.  $S$ has a standard complex Hilbert space $\mathcal{H}_S$.  A two-state clock may not seem like a very good clock, but it is enough to tell us whether time is going forward (the $|+\ran_C$ state) or backwards (the $|-\ran_C$ state).  The action of $\Theta$ on the combined system is
\be
\Theta\Big(|\pm\ran_C\otimes |\psi\ran_S\Big)=|\mp\ran_C\otimes \Theta_S|\psi\ran_S,
\ee
where $\Theta_S$ is an antiunitary operator that implements time reversal just on the system $S$.  To get a physical state which is invariant under $\Theta$ we need to take a superposition:
\be
|\wt{\psi}\ran_{CS}=\frac{1}{\sqrt{2}}\Big(|+\ran_C\otimes |\psi\ran_S+|-\ran_C\otimes \Theta_S|\psi\ran_S\Big).
\ee
If $|\psi\ran_S$ is invariant under $\Theta_S$ then this is a product state, otherwise there is entanglement between the system and the clock.  Now let's say that we would like to do a measurement projecting onto an arbitrary state $|\chi\ran_S$.  Such a measurement in general is not $\Theta$-invariant, but on the total system we can instead measure the $\Theta$-invariant projection
\be
\wt{P}=|+\ran\lan+|_C\otimes |\chi\ran\lan\chi|_S+|-\ran\lan-|_C\otimes |\Theta_S \chi\ran\lan\Theta_S\chi|_S.
\ee
The result of this measurement is
\begin{align}\nonumber
\lan\wt{\psi}|\wt{P}|\wt{\psi}\ran_{CS}&=\frac{1}{2}\Big(|\lan\psi|\chi\ran_S|^2+|\lan\Theta_S\psi|\Theta_S\chi\ran_S|^2\Big)\\
&=|\lan\psi|\chi\ran_S|^2,
\end{align}
which is just the result we would have gotten by ignoring the clock and doing ordinary quantum mechanics on $S$.  For completeness we can also consider the more general case where $\Theta$ has order $m$, meaning that $\Theta^m=1$, in which case the clock needs to have $m$ states.  Labeling these as $|0\ran_C, |1\ran_C\ran,\ldots |m-1\ran_C$, the above equations are replaced by
\begin{align}\nonumber
\Theta\Big(|n\ran_C\otimes |\psi\ran_S\Big)&=|n+1\ran_C\otimes \Theta_S|\psi\ran_S\\\nonumber
|\wt{\psi}\ran_{CS}&=\frac{1}{\sqrt{m}}\sum_{n=0}^{m-1}|n\ran_C\otimes\Theta_S^n|\psi\ran_S\\
\wt{P}&=\sum_{n=0}^{m-1}|n\ran\lan n|_C\otimes |\Theta_S^n\chi\ran\lan\Theta_S^n\chi|,
\end{align}
with the addition in the clock being mod $m$.  Since the states $|\psi\ran_S$ and $|\chi\ran_S$ can be arbitrary complex superpositions, we thus see that all the usual phenenoma arising from such superpositions can be accounted for using only $\Theta$-invariant states and observables.    

The discussion of the previous paragraph may seem to have the feel of an accounting trick: given any quantum system $S$ we can extend it to include a clock $C$ and then ``lift'' the usual rules of quantum mechanics to take place in a real vector space with one more degree of freedom.  This however is misleading: in a closed universe we are not allowed to use any degrees of freedom which are not already present in the system, so if there is no clock around then we cannot perform the above construction.  And moreover even when we do have a clock, due to the finite size of the universe there will be unsuppressable interactions between the system and the clock which prevent the operation $\Theta_S$ from really being a symmetry.  The correct interpretation of our construction is therefore the following: the usual rules of quantum mechanics on the system $S$ with a time-reversal symmetry $\Theta_S$ arise as a limit where the size of the universe $L$ goes to infinity.  To say this more provocatively, at finite $L$ the standard rules of quantum mechanics with a complex Hilbert space are only approximations to a new set of rules acting on a real Hilbert space.  Such corrections are much too small to be observed in practice, but we believe that this phenomenon is of crucial importance for attempts to construct a complete holographic theory of a closed universe since the real structure would need to be manifest in the fundamental description of the system.

As an application of the idea that the total state of the universe must be $\CRT$-invariant, we briefly recall a classic problem in quantum cosmology.  In the early 1980s it was observed by Vilenkin, Hartle, and Hawking that the Euclidean gravity path integral can be used to define a rather nice candidate for the quantum state of the universe \cite{Vilenkin:1982de,Hartle:1983ai}.  There is some ambiguity in this proposal however, and in particular the wave function one obtains has two natural branches - one expanding and one contracting.    More explicitly, in four spacetime dimensions the Euclidean gravity action with positive cosmological constant is
\be
S_E=-\frac{1}{16\pi G}\int d^4x \sqrt{g} (R-12)-\frac{1}{8\pi G}\int d^{3}x\sqrt{\gamma}K.
\ee
Here we have chosen units so that the maximally symmetric solution is a round $\mathbb{S}^4$ of radius one.  In the ``mini-superspace approximation'' we consider only spatial geometries consisting of a round $\mathbb{S}^{3}$ of radius $a$, and we then study the wave function as a function of $a$.  When $0<a<1$ there are two real solutions whose boundary consists of a three-sphere of radius $a$; in coordinates where the full spacetime metric is
\be
ds^2=d\tau^2+\sin^2\tau d\Omega_{3}^2,
\ee  
they consist of the regions where $\tau\in[0,\tau_c]$ with $\tau_c\in[0,\pi]$ and $\sin\tau_c =a$.  Note that when $0<a<1$ this last equation indeed has two solutions in the allowed range, one with $\tau_c<\pi/2$ and one with $\tau_c>\pi/2$ (they become degenerate when $a=1/2$). These solutions are shown in figure \ref{hhsaddlesfig}.  Their actions are given by
\be
S_E=-\frac{\pi}{2G}\Big(1\pm (1-a^2)\sqrt{1-a^2}\Big),
\ee 
with the upper sign corresponding to the case where $\tau_c>1/2$ and the lower sign corresponding to $\tau_c<1/2$.  Thus in the semiclassical approximation the wave function for a universe of radius $a$ has the two branches
\be
\psi_\pm(a)\approx\exp\left[\frac{\pi}{2G}\Big(1\pm (1-a^2)\sqrt{1-a^2}\Big)\right].
\ee
To apply this wave function to cosmology in Lorentzian signature we need to analytically continue to $a\gg 1$, which gives
\begin{align}\nonumber
\psi_\pm(a)&\approx\exp\left[\frac{\pi}{2G}\Big(1\pm i(a^2-1)\sqrt{a^2-1}\Big)\right]\\
&\sim \exp\left[\pm i\frac{\pi}{2G}a^3\right]
\end{align}
where we have arbitrarily chosen the direction around the branch cut to give the indicated sign and the second expression gives the large $a$ asymptotics. In particular we see that the $+$ branch is expanding and the $-$ branch is contracting: these are the two branches mentioned above.  

\bfig
\includegraphics[height=4cm]{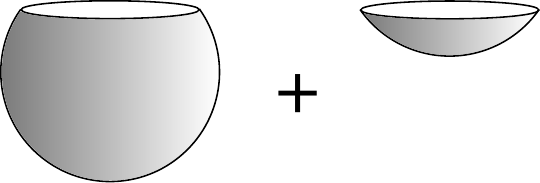}\caption{Summing over two saddles for the wave function of the universe.  After analytic continuation to large spatial geometries these become expanding and contracting branches of the wave function, with the superposition being $\CRT$-invariant.}\label{hhsaddlesfig}
\efig
We thus are naturally confronted with the question of how to choose between these branches.  In \cite{Vilenkin:1982de} it was proposed that one should use only the expanding branch, with the motivation being that we should consider the creation of the universe as a tunneling event, while in \cite{Hartle:1983ai} it was argued that to set the wave function to zero on three-geometries which are not positive-definite it is natural to adopt an integration contour which results in taking a superposition of the two branches with equal coefficients.  We do not view either of these arguments as decisive, in particular the Hartle-Hawking proposal relies on assumptions about small three-geometries which do not seem so compelling since for such geometries one would expect Planck-suppressed corrections to be important.  The point we wish to make however is that the gauging of $\CRT$ gives an ``infrared'' reason for adopting the Hartle-Hawking prescription: it is only when we sum over expanding and contracting branches with equal real coefficients that we get a state which is $\CRT$ invariant!  Another way to think about this is the following: in Euclidean quantum gravity it is natural to sum over all geometries with the desired boundary conditions, and that includes summing over both branches of the wave function of the universe.  

There is a puzzling feature of the idea that a closed universe should have a real vector space which we close this section by discussing.  What if we have a disconnected universe consisting of a closed baby universe and an asymptotically $AdS$ geometry, for example as appears in the final state of figure \ref{conservationfig}.  The usual rule for composing independent systems in quantum mechanics is to take the tensor product, but how can we take the tensor product of a real vector space with complex vector space?  The answer had better be a complex vector space, since after all it seems likely that there are states in AdS/CFT that have a bulk interpretation which includes a baby universe.  We believe the answer is the following: let $\mathcal{H}_A$ be the asymptotically-AdS Hilbert space with a connected spatial slice and $\mathcal{H}_B$ be the (real) baby universe Hilbert space.  Moreover let $|n\ran_A$ be a basis for $\mathcal{H}_A$ and $|i\ran_B$ be a basis for $\mathcal{H}_B$.  Then the allowed (pure) states of the joint system should be all superpositions of the form
\be
|\psi\ran_{AB}=\sum_{ni}C_{ni}|n\ran_A|i\ran_B,
\ee
with $C_{ni}$ allowed to be arbitrary complex numbers such that $\sum_{ni}|C_{ni}|^2=1$.  This is perhaps puzzling since if we compute the reduced state on $B$ we get
\be
\rho_B=\sum_{i,i',n}C_{ni}C_{ni'}^*|i\ran\lan i'|_B,
\ee
which in general is not a real operator, but the point is that for computing expectation values of $\CRT$-invariant observables on $B$ we can get all the same expectation values by instead using
\be
\wt{\rho}_B=\frac{1}{2}\sum_{i,i',n}\left(C_{ni}C_{ni'}^*+C_{ni}^*C_{ni'}\right)|i\ran\lan i'|_B.
\ee
In other words in this context the partial trace should be followed by an average over $\CRT$ in defining the reduced state.

\section{Fermions}\label{fermsec}
So far we have avoided the notoriously tricky question of how spacetime inversions act on spinor fields.  We do not intend to give an exhaustive treatment of the subject, which anyways these days has become an entire industry in the context of topological phases of matter (see \cite{Witten:2015aba} for an introduction accessible to high-energy physicists).  In this section we instead give a somewhat sketchy overview of the subject, focusing on those aspects which are needed to extend our discussion of gauging $\R$ and/or $\C\T$ from section \ref{sumsec} to theories with spinor fields.  Our general approach is quite similar to that of \cite{Witten:2015aba}, although we work in Lorentzian signature instead of Euclidean and are in some places more explicit.  

The study of spinors begins with the realization that the identity component $SO^+(d-1,1)$ of the Lorentz group is not simply-connected for $d\geq 3$.  For $d\geq 4$ we have
\be
\pi_1(SO^+(d-1,1))=\mathbb{Z}_2,
\ee
with the nontrivial element being representable as a loop in the space of rotations that begins with the trivial rotation and smoothly deforms it to a rotation by $2\pi$.  For $d=3$ we instead have
\be
\pi_1(SO^+(2,1))=\mathbb{Z},
\ee
with the generator again being a loop from a $0$-rotation to a $2\pi$-rotation.  When $d=2$ we have $SO^+(1,1)\cong \mathbb{R}$ which is simply-connected, but even there we can go to Euclidean signature to get $SO(2)\cong \mathbb{S}^1$ which has fundamental group $\mathbb{Z}$.  For $d\geq 4$ we can therefore introduce a universal covering group $Spin^+(d-1,1)$, which distinguishes between a trivial rotation and a rotation by $2\pi$ and thus gives a double cover of $SO^+(d-1,1)$.  Spinor fields are precisely those fields which transform in representations of $Spin^+(d-1,1)$ where rotation by $2\pi$ is represented as $-1$; such representations are \textit{projective} when viewed as representations of $SO^+(d-1,1)$, meaning that the group multiplication rule is respected only up to a phase.  For $d=2,3$ there are more possibilities since $Spin^+(d-1,1)$ has nontrivial $\pi_1$, and thus is no longer the universal covering group of $SO^+(d-1,1)$, which leads to the possibility of anyons \cite{Leinaas:1977fm,Wilczek:1981du,Wilczek:1982wy}. In this paper however we will only consider bosonic and fermionic statistics.

Spinor representations of $Spin^+(d-1,1)$ are constructed explicitly by finding a vector $\gamma^\mu$ of $2^{\lfloor d/2\rfloor}\times 2^{\lfloor d/2\rfloor}$ matrices (here $\lfloor x\rfloor$ means the largest integer which is less than or equal to $x$) obeying the Dirac algebra
\be
\{\gamma^\mu,\gamma^\nu\}=2\eta^{\mu\nu}.
\ee
When $d$ is even an explicit representation of this algebra on $d/2$ qubits is given by
\begin{align}\nonumber
\gamma^0&=i\sigma_y\otimes I\otimes\ldots I\\\nonumber
\gamma^1&=\sigma_x\otimes I\otimes\ldots I\\\nonumber
\gamma^{2a}&=\sigma_z\otimes\ldots \otimes \sigma_z\otimes \sigma_x\otimes I\otimes\ldots \otimes I\\
\gamma^{2a+1}&=\sigma_z\otimes\ldots \otimes \sigma_z\otimes \sigma_y\otimes I\otimes\ldots \otimes I,\label{gammas}
\end{align}
with the $\sigma_x/\sigma_y$ appearing in the $(a+1)$st spot in the last two lines and $a=1,2,\ldots (d-2)/2$.  When $d$ is odd a representation is given by taking the $\gamma$-matrices just constructed for dimension $d-1$ and including the extra matrix
\be
\gamma^{d-1}=i^{-(d-3)/2}\gamma^0\ldots \gamma^{d-2}.
\ee
In any dimension the Lorentz generators are then given by
\be
J^{\mu\nu}_{Dirac}=-\frac{i}{4}[\gamma^\mu,\gamma^\nu], 
\ee
in terms of which we have the spinor transformation
\be\label{spinortrans}
U(e^{-i\omega_{\mu\nu}J^{\mu\nu}})\psi(x)U(e^{i\omega_{\mu\nu}J^{\mu\nu}})=e^{i\omega_{\mu\nu}J^{\mu\nu}_{Dirac}}\psi\left(e^{-i\omega_{\mu\nu}J^{\mu\nu}} x\right).
\ee
In particular we have
\be
J^{2a,2a+1}_{Dirac}=I\otimes \ldots I\otimes \frac{\sigma_z}{2}\otimes I\otimes \ldots\otimes I,
\ee
and thus
\be
e^{2\pi i J^{2a,2a+1}_{Dirac}}=-1,
\ee
so a spinor field indeed changes by a sign under a rotation by $2\pi$ and thus lives in a representation of $Spin^+(d-1,1)$ instead of $SO^+(d-1,1)$.  The transformation \eqref{spinortrans} leaves invariant the Dirac Lagrangian
\be\label{DiracL}
\mathcal{L}=-i\ol{\psi}\gamma^\mu\partial_\mu \psi-im\ol{\Psi}\Psi,
\ee
where as usual we have defined
\be
\ol{\psi}=\psi^\dagger \gamma^0
\ee
with $\dagger$ indicating both the adjoint on Hilbert space and the transpose on Dirac indices. 

The basic problem with defining spacetime inversions of spinors is that, unlike for tensor fields, the transformation law \eqref{spinortrans} only determines the behavior of spinors under coordinate transformations that are connected to the identity.  We therefore are lacking any fundamental principle which tells us how they should transform under $\R$ and $\T$.  The conventional approach in particle physics is to require these transformations to preserve the massive Dirac Lagrangian \eqref{DiracL}, which leads to the transformations\footnote{For completeness we'll mention that the $\C$ symmetry that preserves this Lagrangian is given for $d=0,2,3 \, \mathrm{mod}\, 4$ by
\be
U_\C^\dagger \psi(x)U_\C=e^{i\theta_\C}B_\C^\dagger \psi^*(x),
\ee
with $B_\C=\gamma B_\T$ for $d$ even and $B_C=\gamma^3\gamma^5\ldots \gamma^{d-2}$ for $d=3 \, \mathrm{mod}\, 4$.  For $d=1 \, \mathrm{mod} \, 4$ there is no $\C$ symmetry. 
}
\begin{align}\nonumber
U_\R^\dagger \psi(x)U_\R&=e^{i\theta_\R}\gamma \gamma^1 \psi(\R x)\qquad \qquad \qquad d \,\, \mathrm{even}\\
U_\T^\dagger \psi(x)U_\T&=e^{i\theta_\T}\gamma^0 B_\T \psi(\T x)\qquad \qquad d=0,1,2 \,\mathrm{mod}\, 4,
\end{align}
with
\be
\gamma=i^{-(d-2)/2}\gamma^0\ldots \gamma^{d-1}
\ee
and\footnote{These expressions for $B_\T$ are valid for the specific representation \eqref{gammas} of the $\gamma$ matrices.  For another representation $\gamma^{\mu\prime}=U\gamma^\mu U^\dagger$ with $U$ unitary, we should take $B'_\T=U^* B_\T U^\dagger$.  The same transformation applies to $B_\C$ from the previous footnote.}
\be
B_\T=
\begin{cases}
\gamma^3\gamma^5\ldots \gamma^{d-1} & d=0 \,\mathrm{mod} \,4\\
\gamma^3\gamma^5\ldots \gamma^{d-2} & d=1 \,\mathrm{mod} \,4\\
\gamma \gamma^3\gamma^5\ldots \gamma^{d-1} & d=2 \,\mathrm{mod} \,4
\end{cases}.
\ee
Here the phases $e^{i\theta_\R}$ and $e^{i\theta_\T}$ are arbitrary, reflecting the fact that the Dirac Lagrangian preserves fermion number and is thus invariant under 
\be
\psi'=e^{i\theta}\psi.
\ee
Note also that there is no definition of time-reversal that preserves \eqref{DiracL} for $d=3\, \mathrm{mod} \,4$.\footnote{One can however define an $\RT$ symmetry for $d=3\, \mathrm{mod} \,4$ even though $\R$ and $\T$ are not separately symmetries.  Similarly a $\C\R$ symmetry can be defined for $d=1 \, \mathrm{mod}\, 4$ even though there $\C$ and $\R$ are not separately symmetries.  Thus the combination $\CRT$ makes sense in any dimension, as it had better.}    In general however even preserving \eqref{DiracL} is too restrictive, since not all spinors in nature are governed by it.  Indeed we could set $m=0$, which allows for more possibilities (such as dropping the factor of $\gamma$ in the $\R$ transformation), we could impose some kind of reality constraint (as we did in the theory \eqref{IsingL} above), and in even dimensions we could have chiral fermions where we keep only the components of $\psi$ with $\gamma=\pm 1$.  Giving a general treatment of all cases of interest would be quite tedious, so we leave the discussion here and turn to spinors in curved space.

The obvious proposal for defining spinor fields on a Lorentzian manifold $M$ would be to look for matrices $\gamma^\mu(x)$ obeying
\be
\{\gamma^\mu(x),\gamma^\nu(x)\}=2g^{\mu\nu}(x)
\ee
everywhere in $M$, but unfortunately this is only possible on the rather restricted set of ``parallelizeable'' manifolds whose tangent bundle is trivial.  A more promising idea is to define these matrices in patches using the vielbein: in each coordinate patch $U_\alpha$ we define
\be
\gamma^\mu(x)\equiv \gamma^a e^{\phantom{a}\mu}_a(x),
\ee
where as in section $\ref{sumsec}$ $\{e^{\phantom{a}\mu}_a(x)\}$ are a local orthonormal frame in $U_\alpha$ obeying
\be
e^{\phantom{a}\mu}_a(x)e^{\phantom{a}\nu}_b(x)g_{\mu\nu}(x)=\eta_{ab}.
\ee
Covariant derivatives of spinors in each patch are then defined using the \textit{spin connection}
\be
\nabla_\mu\psi=\left(\partial_\mu+\frac{i}{2}J_{Dirac}^{ab}\,g_{\rho\sigma}e_a^{\phantom{a}\rho}\nabla_\mu e_b^{\phantom{b}\sigma}\right)\psi,
\ee
with more general representations being handled by replacing $J_{Dirac}^{ab}$ by the appropriate Lorentz generator for each representation.  The trouble begins when we consider how to glue together different patches: naively we should do this using the matrices $(D_{\alpha\beta})^a_{\phantom{a}b}$ defined in section \ref{sumsec}, but there is a problem in that the $D_{\alpha\beta}$ matrices live in $O(d-1,1)$ while spinors live in representations of $Spin^+(d-1,1)$.  These groups differ in two ways:
\bi
\item[(1)] $O(d-1,1)$ has four connected components, while $Spin^+(d-1,1)$ has only one.
\item[(2)] $Spin^+(d-1,1)$ is a double cover of the identity component $SO^+(d-1,1)$ of $O(d-1,1)$.
\ei

We can provisionally dispense with the first problem by not gauging any of $\R$, $\C\T$, or $\C\R\T$, in which case we only need to consider Lorentzian manifolds that are both oriented and time-oriented.  To deal with the second problem we need a way to consistently lift the $D_{\alpha\beta}$ matrices to elements of $Spin^+(d-1,1)$.  In other words in each double overlap $U_\alpha\cap U_\beta$ we need a ``lifting map'' 
\be
L_{\alpha\beta}:SO^+(d-1,1)\to Spin^+(d-1,1)
\ee
such that
\be\label{Lproj}
\pi\circ L_{\alpha\beta}=Id_{SO^+(d-1,1)}
\ee 
where $\pi:Spin^+(d-1,1)\to SO^+(d-1,1)$ is the quotient map.  Spinors are then glued using the rule (here we are identifying groups with their defining representations to reduce notation)
\be
\psi_\alpha(x_\alpha)=L_{\alpha\beta}(D_{\alpha\beta}(x_\beta))\psi_\beta(x_\beta).
\ee
For this rule to be self-consistent however we need it to obey a triple overlap rule: for all $x\in U_\alpha\cap U_\beta\cap U_\gamma$ we need to have
\be\label{Ltriple}
L_{\alpha\beta}(D_{\alpha\beta}(x))L_{\beta\gamma}(D_{\beta\gamma}(x))L_{\gamma\alpha}(D_{\gamma\alpha}(x))=1.
\ee
A choice of $L_{\alpha\beta}$ obeying this rule is called a \textit{spin structure on $M$}, and a manifold $M$ equipped with such a structure is called a \textit{spin manifold}.  The triple overlap rule will not be satisfied by an arbitrary choice of $L_{\alpha\beta}$ obeying \eqref{Lproj}, and indeed not all manifolds admit a spin structure (the ones that do however are a much bigger set than the ones which are parallelizeable).  In general the same manifold can admit more than one spin structure, and in fact these are essentially classified by the $\pm 1$ holonomies of spinors around the loops in $M$.  We can therefore view spin structures as background gauge fields for fermion parity, as in section \ref{intexsec}.  Since fermion parity is a rotation by $2\pi$, in a gravitational theory it should always be gauged and therefore in the path integral formulation we should always sum over spin structures on $M$.\footnote{If the fermions in question are charged under an internal gauge symmetry then there is a more general kind of possible background, called a $\mathrm{Spin_C}$ structure, where the $L_{\alpha\beta}$ fail the triple overlap condition but in a way that can be fixed by turning on a nontrivial background for the internal gauge field.  Backgrounds of this type should also be summed over.}  

Now we can consider gauging some version of $\R$ and/or $\T$.  Unfortunately there is not much that can be said in general, since as we just discussed the actions of $\R$ and $\T$ on spinors are quite theory-dependent.  Two fairly broad sets of cases which can be understood in fairly simple terms are those for which
\be
\R^2=1,
\ee
in which case the disconnected generalization of $Spin^+(d-1,1)$ is called $Pin_+(d-1,1)$, and those for which
\be
\R^2=(-1)^F,
\ee
in which case the disconnected generalization of $Spin^+(d-1,1)$ is called $Pin_-(d-1,1)$.  Since we always have $(\CRT)^2=1$, this means that in $Pin_+(d-1,1)$ we have
\be
(\C\T)^2=1
\ee
while in $Pin_-(d-1,1)$ we have
\be
(\C\T)^2=(-1)^F.
\ee
 $Pin_\pm(d-1,1)$ each have four connected components, and each gives a double cover of $O(d-1,1)$.  We therefore again need to lift the $O(d-1,1)$-valued $D_{\alpha\beta}$ matrices to $Pin_{\pm}(d-1,1)$-valued matrices using lifting maps $L_{\alpha\beta}:O(d-1,1)\to Pin_{\pm}(d-1,1)$ that obey projection and triple overlap conditions analogous to \eqref{Lproj} and \eqref{Ltriple}; this gives what is called a $Pin_{\pm}(d-1,1)$ structure on $M$.  Restricting for simplicity to the case where $\CRT$ is gauged, in which case gauging $\R$ means that $\C\T$ is automatically gauged, in this language gauging $\R/\C\T$ means that we should sum over manifolds with structure group $O(d-1,1)$ and also sum over $Pin_{\pm}(d-1,1)$ structures on those manifolds.  Theories with more complicated $\R$ and $\T$ symmetries are probably best dealt with on a case-by-case basis.

\section{Discussion}\label{conc}
In this final section we make some comments about how the things we have discussed in this paper relate to some other recent developments in theoretical physics.

\subsection{Completeness for spacetime inversions}\label{comsec}
\bfig
\includegraphics[height=5cm]{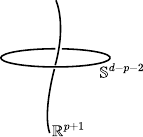}
\caption{Linking the world-volume of a $p$-brane in $d$ spacetime dimensions with an $\mathbb{S}^{d-p-2}$ on which it can source topological charge.}\label{linkingfig}
\efig
A conjecture about symmetry in quantum gravity that is often discussed together with the ``no global symmmetries'' conjecture is the ``completeness hypothesis'', which says that there should be dynamical objects carrying all possible gauge charges \cite{Polchinski:2003bq,Banks:2010zn}.  This idea applies to any kind of conserved charge which can be defined from the gauge field, for example in the $U(1)$ case it includes both electric and magnetic charge.  A nice formulation of this idea is the ``cobordism conjecture'' of \cite{McNamara:2019rup}, which roughly speaking says that for any topological charge measurable at infinity there must always be some object which can source it.   For example let's say we have an internal (zero-form) gauge symmetry $G$.  The objects which are ``electrically'' charged under this gauge symmetry are pointlike, so we can call them $0$-branes.  There can also however be branes of various dimensions carrying various kinds of topological charge associated to this symmetry.  More precisely, in order to have a $p$-brane which carries a topological charge under $G$, we need to be able to define this charge on the $\mathbb{S}^{d-p-2}$ which links with the worldvolume of this brane in spacetime (see figure \ref{linkingfig}). In order to have a topological charge on this $\mathbb{S}^{d-p-2}$, we need the equatorial gauge transformation which glues its northern and southern hemispheres to live in a nontrivial element $\sigma\in\pi_{d-p-3}(G,g_0)$ where $g_0$ is some base point.  When $G$ is disconnected these elements do not all correspond to physically distinct charges: a gauge transformation in the northern or southern hemisphere multiples $\sigma$ on the left or right by a gauge transformation which is homotopic to a constant.  We can use this freedom to set $g_0=e$, after which the only remaining gauge equivalence is conjugation of $\sigma$ by a constant element of $G$.  Completeness thus says that for each nontrivial element $\sigma\in\pi_n(G,e)$ modulo conjugation by a constant gauge transformation, there should be a dynamical $p$-brane with
\be
p=d-n-3
\ee
that sources a gauge field on $\mathbb{S}^{d-p-2}$ with equatorial gluing in $\sigma$. For example we can consider $d=4$: if $\pi_0(G,e)$ is nontrivial then there are strings around which the gauge field has a holonomy by some $g\in G$ which is not in the identity component, while if $\pi_1(G,e)$ is nontrivial then there are pointlike magnetic monopoles.   If $G$ is discrete, then in any dimension the only needed objects are $(d-3)$-branes around which there is $G$-holonomy \cite{Krauss:1988zc}, and these branes are labeled by conjugacy classes of $G$.  

In the context of AdS/CFT there is a nice way to think about completeness for topological charges.  Given an internal gauge symmetry in the bulk with gauge group $G$, there will be a global $G$-symmetry in the boundary CFT \cite{Witten:1998qj,Harlow:2018tng}.  We therefore can turn on a background gauge field for $G$ in the boundary.  From the bulk point of view this is imposing a boundary condition that there is some nonvanishing topological charge measureable at infinity.  In particular if $G$ is discrete then we can take the boundary to have topology $\mathbb{R}^{d-2}\times \mathbb{S}^1$, with some nontrivial $G$-holonomy around the $\mathbb{S}^1$ which is classified by conjugacy classes of $G$ as we discussed in section \ref{intsec}. We can then ask what happens in the bulk in the presence of such a boundary condition?  The most natural way to view $\mathbb{R}^{d-2}\times \mathbb{S}^1$ as the boundary of a bulk geometry is to view the $\mathbb{S}^1$ as the boundary of a disk, but this would require the $G$-holonomy to be trivial.  To avoid this we need to insert an object at the center of the disk which has nontrivial $G$-holonomy, and this object is precisely the $(d-3)$-brane required by completeness.  In the language of \cite{McNamara:2019rup}, $\mathbb{S}^1\times\mathbb{R}^{d-2}$ with $G$-holonomy is not null-cobordant so there must be some object that sources it. This object could be a brane of sub-Planckian tension with a world-volume action, or it could be a black object with a horizon.  In the latter case we can consider the $AdS_d$ black brane metric
\be\label{bbrane}
ds^2=r^2\left[-\left(1-\frac{\alpha}{r^{d-1}}\right)dt^2+d\phi^2+d\vec{x}^2_{d-3}\right]+\frac{dr^2}{r^2\left(1-\frac{\alpha}{r^{d-1}}\right)},
\ee
where $d\vec{x}^2_{d-3}$ indicates the Euclidean metric on $\mathbb{R}^{d-3}$ and $\alpha$ is proportional to the boundary energy density.  The full geometry describes a wormhole connecting two boundaries with topology $\mathbb{R}^{d-2}\times \mathbb{S}^1$, and the $\phi$ circle does not contract away from the singularities at $r=0$ so it is allowed to have nontrivial $G$-holonomy.  The one-sided microstates of this wormhole are black $(d-3)$-branes with nontrivial $G$-holonomy.

What about completeness for spacetime inversions?  It is clear that any nonconstant local field will transform nontrivially under $\R$, $\T$, and $\RT$, so in most theories of gravity the ``electric'' charges are manifestly present.  Even in low-dimensional theories that do not have matter fields or gravitons, as long as black holes exist these already give objects that are ``electrically'' charged under $\R$, $\T$, and $\RT$.  More interesting is to consider the possibility of topological charge for $\R$, $\T$, and $\RT$.  What this means is the following: should we require the existence of objects around which there is nontrivial $\R$, $\T$, or $\RT$ holonomy?  

\bfig
\includegraphics[height=5cm]{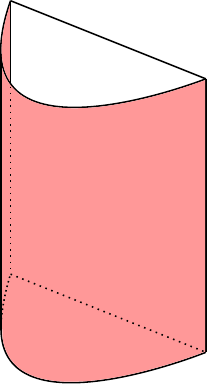}
\caption{An end-of-the-world brane, shaded red, cutting off an asymptotically-$AdS_3$ geometry whose boundary geometry is $\mathbb{R}\times I$.}\label{eowfig}
\efig
We first consider the case of branes about which there is holonomy for $\R$. This possibility was discussed extensively in \cite{McNamara:2022lrw}.\footnote{We thank Jake McNamara for useful comments on a previous version of this subsection.}  We need to specify the direction in which $\R$ acts as we move around the brane, and there are three possibilities:  1) in the direction of the holonomy circle, 2) in the radial direction, or 3) along the brane directions.  In case 1) the quotient turns the holonomy circle into a closed line interval with two boundaries, as explained in section \ref{Rsubsec}.  We should therefore look for states of the boundary CFT on the spatial topology $I\times \mathbb{R}^{d-3}$, where $I$ is a closed line interval.  These cannot be thought of as being sourced by a $(d-3)$ brane, we instead should think of them as describing an end-of-the-world brane with $d-2$ spatial dimensions (see figure \ref{eowfig}).  Completeness in case 1) therefore predicts the existence of such an end-of-the-world brane \cite{McNamara:2022lrw}.  Case 2) is somewhat unusual, in that a reflection in the radial direction does not preserve the notion of asymptotic infinity.  We therefore should not think of case 2) as requiring the existence of an object.  For case 3), we need to be in high enough dimensions that there is a direction orthogonal to the time direction, the radial direction, and the holonomy circle.  In other words we need $d\geq 4$.  We can construct a boundary spatial geometry with such a holonomy by starting with $\mathbb{S}^1\times \mathbb{R}^{d-3}$ and then performing the identification
\be
(\phi,x,\vec{y})\sim(\phi+\pi,-x,\vec{y}),
\ee
where $\vec{y}$ lives in $\mathbb{R}^{d-4}$. This boundary spatial topology is $M_2\times \mathbb{R}^{d-4}$, where $M_2$ is the noncompact M\"obius strip.  We can then consider what kind of bulk geometry can fill in this space.  The only possibility is to try to view the circle as the boundary of a disk, giving a bulk spatial geometry constructed by the identification
\be
(r,\phi,x,\vec{y})\sim(r,\phi+\pi,-x,\vec{y}).
\ee
The fixed points of this identification can be viewed as the location of the spatial volume of a brane which sources this $\R$-holonomy.  Here we meet a surprise \cite{McNamara:2022lrw}: the fixed points lie in the locus $r=0,\, x=0$, which describe a $(d-4)$-dimensional spatial surface.  Thus a brane with $\R$ holonomy of type 3) should be a $(d-4)$-brane instead of a $(d-3)$-brane!  In \cite{McNamara:2022lrw} these $(d-4)$-branes were called ``I-folds''.  On the other hand it seems that, although we cannot make a localized $(d-3)$-brane which sources $\R$-holonomy, we \textit{can} make a black $(d-3)$-brane which does.  The construction is simple: beginning with the black brane geometry \eqref{bbrane}, we split $\vec{x}$ into $(x,\vec{y})$ and then make the spatial identification
\be
(X,\phi,x,\vec{y})\sim(X,\phi+\pi,-x,\vec{y}),
\ee
where $X$ is the Kruskal spatial coordinate that goes through the wormhole from one side to the other.  This identification has no fixed points and produces a wormhole between two boundaries whose spatial topology is each $M_2\times \mathbb{R}^{d-4}$.  This is also the topology of the horizon.  This geometry has spatial translation symmetry in $d-3$ directions, and thus its one-side microstates should be viewed as black $(d-3)$-branes around which we have $\R$ holonomy of type 3).

Turning now to the case of $\T$, are there branes with the property that if you circle them then time reverses direction? Proceeding as in the previous paragraph, we can consider a boundary spacetime geometry constructed by beginning with $\mathbb{R}\times \mathbb{S}^1\times \mathbb{R}^{d-3}$ and then performing the identification
\be
(t,\phi,\vec{x})\sim(-t,\phi+\pi,\vec{x}).
\ee
This spacetime geometry has topology $L_2\times \mathbb{R}^{d-3}$, where $L_2$ is the Lorentzian M\"obius strip.  Proceeding as before we can try to view this as the boundary of a bulk spacetime where the $\phi$ circle contracts:
\be
(t,r,\phi,\vec{x})\sim(-t,r,\phi+\pi,\vec{x}).
\ee
This geometry has a singularity at $r=0,\,t=0$, which we can view as a spacelike $(d-4)$-brane.  It therefore should not be viewed as an object, it is more like an instanton with spatial extent. On the other hand we \textit{can} again make a black $(d-3)$-brane which has nontrivial $\T$-holonomy.  The construction is analogous to what we did for $\R$: beginning with the black brane \eqref{bbrane}, in Kruskal coordinates we make the identification
\be
(T,X,\phi,\vec{x})\sim(-T,X,\phi+\pi,\vec{x}).
\ee 
This has no fixed points, and produces a wormhole connecting two asymptotic boundaries with spacetime geometry $L_2\times \mathbb{R}^{d-3}$.  An observer near either of the wormhole exits would view themselves as being in the vicinity of a $(d-3)$-brane around which there is a $\T$-holonomy, and the one-sided microstates of this geometry (which live in the Hilbert space of the CFT constructed on $L_2\times \mathbb{R}^{d-3}$ as discussed in section \ref{tback1}) give concrete examples of such branes.

\subsection{Renormalizeable gravity and random matrix ensembles}  
In recent years it has been understood that in low spacetime dimension there can be renormalizeable theories of gravity with enough structure to model interesting features of higher-dimensional gravity, with the most well-known example being Jackiw-Teitelboim gravity \cite{Jackiw:1984je,Teitelboim:1983ux,Almheiri:2014cka,Maldacena:2016upp,Jensen:2016pah,Engelsoy:2016xyb,Kitaev:2017awl,Harlow:2018tqv}.  In \cite{Harlow:2020bee} it was emphasized however that the ban on global symmetries does not apply in such theories, so the spacetime inversion structure can be more varied.  In this subsection we will discuss some of the possibilities.  

The simplest approach to renormalizeable theories of gravity is canonical quantization, which in pure JT gravity was studied in \cite{Harlow:2018tqv,Yang:2018gdb}.  The result is that pure JT gravity canonically quantized on connected spacetimes with two asymptotically-$AdS_2$ boundaries is equal to the quantum mechanics of a particle moving in an exponential potential.  This theory is $\T$-invariant, but it is not clear whether we should view $\T$ as a gauge or global symmetry: the presence of the boundaries means that we do not need to project to $\T$-invariant states, and there are no connected Lorentzian geometries with these boundary conditions that are not time-orientable. To distinguish between $\T$ being gauge or global we therefore need to introduce additional closed universes, and then specify whether or not their states need to be $\T$-invariant and whether or not they can carry the Lorentzian M\"obius strip topology.  In canonical quantization either choice is valid, consistent with \cite{Harlow:2020bee}.\footnote{This discussion becomes simpler if we add matter fields, since the canonically-quantized theory remains renormalizeable in the presence of matter fields and there is no obstruction to those matter fields having internal global symmetries.}

The situation becomes more interesting once we do path integral quantization: in ordinary field theories this is of course equivalent to canonical quantization, but in quantum gravity they are different \cite{Harlow:2020bee}.  Indeed in pure JT gravity the path integral approach does not define a quantum theory in the usual sense at all: instead it computes an average over an \textit{ensemble} of quantum systems \cite{Saad:2019lba}.  One way to think of the difference between the two is that the path integral naturally sums over processes which change the topology of space, so the additional closed universes mentioned in the previous paragraph become unavoidable.  In this interpretation we can again ask how to think about gauge and global symmetries in the gravitational theory, which now clearly are different due to the presence of closed universes.  A nice answer was proposed in \cite{Hsin:2020mfa}: gauge symmetries of the gravitational theory are symmetries of each member of the ensemble, while global symmetries are symmetries of the ensemble that are broken for each individual sample.  This idea is compelling, but the argument for it is somewhat subtle so we here present a somewhat modified version of it.

\bfig
\includegraphics[height=5cm]{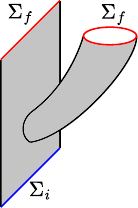}
\caption{Bulk global symmetry conservation in the presence of topology change: the initial and final Cauchy slices are homologous via the grey region, so the charge on $\Sigma_i$ is equal to the charge on $\Sigma_f$.}\label{conservationfig}
\efig
\bfig
\includegraphics[height=7cm]{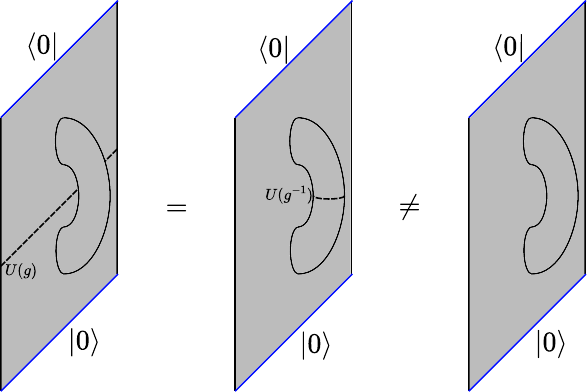}
\caption{Violation of a boundary global symmetry due to topology change \cite{Hsin:2020mfa,Chen:2020ojn}.  Here $|0\ran$ is a state which is invariant under the symmetry $U(g)$, but applying $U(g)$ only to the boundary component of the state after a topology change gives a nontrivial transformation law.}\label{violationfig}
\efig
Indeed let's imagine we have a renormalizeable gravitational theory with a global symmetry, which for convenience we can take to be a continuous $U(1)$ symmetry with Noether current $J$ such that
\be
d\star J=0. 
\ee
By Stokes' theorem we see that the integral of $\star J$ must be the same over any two hypersurfaces that are homologous to each other:
\be
\int_{\Sigma_f}\star J-\int_{\Sigma_i}\star J=\int_{M_{if}}d \star J=0.
\ee
An example of this conservation in the presence of topology change is shown in figure \ref{conservationfig}.  In what sense then can we say that the symmetry is violated?  From a purely bulk point of view it is clear that it isn't, so we need to introduce some sort of nontrivial interpretation of this calculation to conclude the symmetry is violated.  The essential idea is that we need to assume the existence of a holographic dual description, after which we can then conclude that a global symmetry of the bulk theory must also be a global symmetry of the boundary theory (see section 4 of \cite{Harlow:2018tng}). This would then imply that the symmetry should be conserved including \textit{only} the boundary components of $\Sigma_i$ and $\Sigma_f$, but figure \ref{conservationfig} makes it clear that this need not be the case since global charge can end up in a ``baby universe'' part of $\Sigma_f$.  We give a concrete illustration of such boundary global symmetry violation in figure \ref{violationfig}.  On the other hand for a bulk gauge symmetry the boundary charge is indeed conserved even in the presence of topology change: one way to see this is that by Gauss's law the $U(g)$ operator in figure \ref{violationfig} really only has support at its boundaries, and these can be freely deformed past the handle in figure \ref{violationfig}.  Another way to see this is that the $U(g^{-1})$ operator acts on a closed universe, and in a closed universe there is no gauge charge.  The natural interpretation of these results in an ensemble context is that proposed in \cite{Hsin:2020mfa}: figure \ref{conservationfig} shows that a bulk global symmetry is a good symmetry of the ensemble, but figure \ref{violationfig} shows that the symmetry is violated in individual instances.\footnote{Another potential interpretation is that of \cite{Marolf:2020xie}, which rejects the boundary interpretation entirely and views the path integral including topology change as defining a nontrivial inner product between bulk states.  In this context there is no global symmetry violation, but there is also no holography.}

Turning now to spacetime inversion symmetries, for $\T$ symmetry in pure JT gravity the key points were already made in \cite{Stanford:2019vob} so we simply review them here in our language (see also \cite{Yan:2023rjh} for a related discussion in 3D gravity).  Treating $\T$ as a global symmetry in the bulk means that in computing the partition function we should only sum over oriented Euclidean topologies.  This was the problem studied in \cite{Saad:2019lba}, and the appropriate boundary ensemble is the $GUE$ ensemble of arbitrary hermitian matrices (which usually do not have $\T$ symmetry). On the other hand this ensemble \textit{does} have $\T$ symmetry, just as suggested in \cite{Hsin:2020mfa}.  In \cite{Stanford:2019vob} it was then observed that if we now gauge $\T$ in the bulk, in computing the partition function we should include both oriented and unoriented geometries.  It was then shown that this is dual to matrix ensembles of $GOE$ or $GSE$ type, which indeed have the property that they only sum over Hamiltonians which have $\T$ symmetry just as suggested in \cite{Hsin:2020mfa}.

It is natural to ask to what extent these results for JT gravity can be extended to higher dimensions.  The basic problem is that (with the exception of pure gravity in $2+1$ dimensions), gravity in higher-dimensions is not renormalizeable and so there cannot really be a well-defined ensemble interpretation of the path integral since the path integral itself is not well-defined.  What seems to be true in general is that the gravitational path integral can be understood as computing coarse-grained results that are valid for any particular holographic theory with that low-energy action in the bulk.  In such a context calculations such as that shown in figure \ref{violationfig} can be understood as giving a lower bound on global symmetry violation that is of order $e^{-1/G}$ \cite{Chen:2020ojn,Hsin:2020mfa,Bah:2022uyz}, but typically there will be larger effects due to terms in the effective Lagrangian arising from integrating out short-distance physics (i.e. such as strings or Planck-sized black holes) which lead to violations that are only suppressed by powers of the Planck mass.  We therefore should expect that only gauged spacetime inversions are symmetries even at the perturbative level, which is the point of view we have taken in the rest of this paper.

\subsection{Fractionalization of $\R$ and $\C\T$}
One of the more interesting developments in condensed matter physics in recent years, motivated by the fractional quantum hall effect, is the idea of \textbf{fractionalization}.  What this means is that a theory with a global symmetry group $G$ can have low-energy particle-like excitations (``quasiparticles'') that transform in projective representations of $G$.  For example $G$ could be $U(1)$, with units for charge chosen so that the fundamental representation has charge one, and yet there could be quasiparticles of charge $1/3$.  The way this can happen is that the quasiparticles can carry an additional gauge charge, in which case they are created at the endpoints of line operators rather by gauge-invariant local operators, and it is only the latter that must transform in genuine (non-projective) representations of $G$.  A standard example of this is in QCD, where the quarks/antiquarks have electric charges $\pm 1/3$ or $\pm 2/3$ but all gauge-invariant local operators have integer charge (here  we are not including a Maxwell field so electric charge generates a global symmetry).  Of course in QCD the quarks are confined, and so don't really give good quasiparticles, but we can imagine turning on a deformation that deconfines them.  For example we could turn on an AdS background and then go to the regime where $\Lambda_{QCD}\ll 1/\ell_{AdS}$ (see \cite{Aharony:2012jf} and section 3.1 of \cite{Harlow:2018tng}).  Another example of this phenomenon is anyons, which can (and probably should) be viewed as arising from the fractionalization of angular momentum.

In the context of this paper fractionalization arises in the situation where $\CRT$ is gauged but $\R$ and $\C\T$ are global.  Of course we do not expect this to be possible exactly in holographic theories since these have no global symmetries, but the breaking of global symmetries can be quite small in many regimes of interest so in this section we ignore it (or else we work in the ``average theory'' in renormalizeable cases where this makes sense).  In the Euclidean path integral this is the situation where we sum over only oriented geometries, which arguably is the most familiar setting for Euclidean gravity.  For simplicity focusing on theories where $\R$ and $\C\T$ square to one,  before gauging $\CRT$ these generate a $\mathbb{Z}_2\times\mathbb{Z}_2$ global symmetry.  Gauging $\CRT$ means that we should pass to the quotient group $\mathbb{Z}_2$, whose trivial element is the class $\{1, \CRT\}$ and whose nontrivial element is the class $\{\R,\C\T\}$.  Thus any operator which is invariant under $\CRT$ must be acted on in the same way by $\R$ and $\C\T$.  This is perhaps a surprising conclusion, since in daily life we certainly do not expect $\R$ and $\C\T$ to always act in the same way on the particles around us.  The resolution of this puzzle is of course that $\R$ and $\C\T$ are fractionalized.  For example given a free real scalar on $\mathbb{R}^4$ we can consider at $t=0$ the product
\be
\dot{\phi}(0,x,0,0)\dot{\phi}(0,-x,0,0).
\ee
This is a $\CRT$-invariant line operator (the line here is a $\CRT$ Wilson line, which is topological so we don't need to specify its support), but $\R$ and $\C\T$ act differently on its endpoints since $\R$ exchanges them while $\C\T$ acts as a minus sign.  We therefore can say that the $\mathbb{Z}_2$ global symmetry group generated by $\{\R,\C\T\}$ is fractionalized, with quasiparticle excitations that transform distinctly under $\R$ and $\C\T$ as we'd expect from everyday experience.

\paragraph{Acknowledgments:}  We thank Netta Engelhardt, Ted Jacobson, Zohar Komargodski, Ho Tat Lam, Alex Maloney, Juan Maldacena, Don Marolf, Jake McNamara, Greg Moore, Hirosi Ooguri, Shinsei Ryu, Nathan Seiberg, Shu-heng Shao, Jon Sorce, Leonard Susskind, Tadashi Takayanagi, Cumrun Vafa, and Edward Witten for useful discussions.  DH is supported by the Sloan Foundation as a Sloan Fellow, the Packard Foundation as a Packard Fellow, the Air Force Office of Scientific Research under the award number FA9550-19-1-0360, the US Department of Energy under grants DE-SC0012567 and DE-SC0020360, and the MIT department of physics. 
TN is supported by MEXT KAKENHI Grant-in-Aid for Transformative Research Areas A
``Extreme Universe" (22H05248) and JSPS KAKENHI
Grant-in-Aid for Early-Career Scientists (23K13094)

\appendix

\section{Antilinear operators}\label{alapp}
In this appendix we give a quick review of antilinear and antiunitary operators, for the convenience of readers who have not recently considered them.  First recall that on a Hilbert space $\mathcal{H}$ a linear operator $L$ is one which for all $a,b\in \mathbb{C}$ and $|\psi\ran, |\phi\ran\in \mathcal{H}$ obeys
\begin{align}\nonumber
L\Big(a|\psi\ran+b|\phi\ran\Big)&=aL|\psi\ran+bL|\phi\ran\\
&\equiv a|L\psi\ran+b|L\phi\ran,
\end{align}
where in the second line we introduce the convenient notation $|L\psi\ran$ for the action of $L$ on $|\psi\ran$.  The adjoint $L^\dagger$ of $L$ is defined by requiring that for all $|\psi\ran, |\phi\ran\in \mathcal{H}$ we have
\be
\lan \phi|L^\dagger \psi\ran=\lan L\phi|\psi\ran.
\ee
In Dirac notation the two sides of this equation are conflated and both written as $\lan\phi|L^\dagger|\psi\ran$, but we will see momentarily that this is a bad idea for antilinear operators so we instead adopt the following rule: by definition for any operator $O$, linear or antilinear, we have
\be
\lan\phi|O|\psi\ran\equiv \lan \phi|O\psi\ran.
\ee
A unitary operator $U$ is a linear operator for which 
\be
\lan U\phi| U\psi\ran=\lan \phi|\psi\ran
\ee
for all $|\psi\ran, |\phi\ran\in \mathcal{H}$.  

For an antilinear operator $A$ on $\mathcal{H}$ we instead require that
\be
A\Big(a|\psi\ran+b|\phi\ran\Big)=a^*|A\psi\ran+b^*|A\phi\ran,
\ee
and the adjoint is defined so that
\be\label{alad}
\lan\phi|A^\dagger\psi\ran=\lan\psi|A\phi\ran.
\ee
An antiunitary operator $\Theta$ is an antilinear operator obeying 
\be
\lan\Theta\phi|\Theta\psi\ran=\lan\psi|\phi\ran.
\ee

We emphasize that one has to be quite careful about the order of operations in manipulating expressions involving antilinear operators.  For example if $\Theta$ is antiunitary then we have
\be
\label{outeral}
\Theta|\chi\ran\lan\chi|\Theta^\dagger=|\Theta\chi\ran\lan\Theta\chi|
\ee
provided that we interpret the left hand side as the multiplication of three operators, but if we try to derive this by computing the the matrix elements of the left-hand side between two states in the usual Dirac way we run into trouble:
\begin{align}
\lan \phi|\Big(\Theta|\chi\ran \lan \chi|\Theta^\dagger\Big)|\psi\ran=_?\lan \phi|\Theta|\chi\ran \times \lan \chi|\Theta^\dagger|\psi\ran=\lan \phi|\Theta\chi\ran \lan \psi|\Theta \chi\ran,
\end{align}
which is not compatible with the right-hand side of \eqref{outeral}.  A correct derivation is the following:
\begin{align}\nonumber
\lan\phi|\Big(\Theta|\chi\ran\lan\chi| \Theta^\dagger|\psi\ran\Big)&=\lan\chi|\Theta^\dagger\psi\ran^*\lan\phi|\Theta\chi\ran\\\nonumber
&=\lan\psi|\Theta\chi\ran^*\lan\phi|\Theta\chi\ran\\
&=\lan\phi|\Theta\chi\ran\lan\Theta\chi|\psi\ran.
\end{align}
The complex conjugate in the first line arises because we need to move $\lan\chi|\Theta^\dagger\psi\ran$ past $\Theta$ before we act with $\lan \phi|$. Similarly if we have a state $|\Omega\ran$ which is invariant under the adjoint of an antiunitary transformation $\Theta$, then we have the Ward identity
\begin{align}\nonumber
\lan \Omega|O_1\ldots O_n|\Omega\ran&=\lan\Omega|\Theta\Theta^\dagger O_1\Theta\ldots \Theta^\dagger O_n\Theta \Theta^\dagger|\Omega\ran\\\nonumber
&=\lan \Omega|\Theta O_1'\ldots O_n'\Omega\ran\\\nonumber
&=\lan O_1'\ldots O_n'\Omega|\Theta^\dagger\Omega\ran\\\nonumber
&=\lan O_1'\ldots O_n'\Omega|\Omega\ran\\
&=\lan \Omega|O_n^{\prime\dagger}\ldots O_1^{\prime\dagger}|\Omega\ran,\label{wardapp}
\end{align}
where we have defined
\be
O'=\Theta^\dagger O \Theta.
\ee

We will not use it much in this paper, but it is worth mentioning that the spectral theory of antiunitary operators is somewhat different from that of unitary operators.  Indeed given a unitary operator $U$ with a discrete spectrum (this assumption is purely for convenience, the continuous case is treated analogously), by the spectral theorem we can always find an orthonormal basis $|a\ran$ such that for all $a$ we have
\be
U|a\ran=e^{i\theta_a}|a\ran.
\ee
For an antiunitary operator $\Theta$ (with a discrete spectrum) the analogous statement is a bit more complicated \cite{wigner1960normal}: there is an orthonormal basis consisting of a collection of states $\{|i\ran\}$ and a collection of states $\{|a,\pm\ran\}$ such that we have
\begin{align}\nonumber
\Theta|i\ran&=|i\ran\\
\Theta|a,\pm\ran&=e^{\mp i\theta_a/2}|a,\mp\ran,
\end{align}
with the $0<\theta_a \leq \pi$.
The origin of this basis is simple to understand: the $|i\ran$ states are the eigenstates of the unitary operator $\Theta^2$ with eigenvalue one, the $|a,+\ran$ states are eigenstates of $\Theta^2$ with eigenvalue $e^{i\theta_a}$, and the $|a,-\ran$ states are eigenstates of $\Theta^2$ with eigenvalue $e^{-i\theta_a}$. Applying the spectral theorem to $\Theta^2$ we see that these together must give an orthonormal basis.  In this language the set of states which are invariant under $\Theta$ is the set of superpositions of the $|i\ran$ states with real coefficients.

\section{Free partition functions}\label{freeapp}
In this appendix we give (standard, see e.g. \cite{Ginsparg:1988ui}) expressions for the partition functions of free fields in $1+1$ dimensions, emphasizing the interpretations of the various sectors in terms of gauging discrete symmetries.

  For a free scalar in $1+1$ dimensions with target space circumference $2\pi R_\phi$ and spatial circumference $L$, the untwisted partition function \eqref{Z1bos} is given by
\be
Z_1=Z_0 \hat{Z}_1,
\ee
where
\begin{align}\nonumber
\hat{Z}_1&=\prod_{n=1}^\infty\frac{e^{\frac{-2\pi\beta n}{L}}}{\left(1-e^{2\pi i \tau n}\right)\left(1-e^{-2\pi i \ol{\tau} n}\right)}\\
&=\frac{1}{|\eta(\tau)|^2}
\end{align}
is the partition function excluding the zero modes and 
\begin{align}\nonumber
Z_0&=\sum_{n,m=-\infty}^\infty e^{-\frac{1}{2}\mathrm{Im}(\tau)\left(\frac{n^2}{R_\phi^2}+4\pi^2 R_\phi^2 m^2\right)-2\pi i \mathrm{Re}(\tau)nm}\\
&=\sqrt{\frac{2\pi R_\phi^2}{\mathrm{Im} (\tau)}}\sum_{p,m=-\infty}^\infty e^{-\frac{2\pi R_\phi^2}{\mathrm{Im}(\tau)}|p+m\tau|^2}
\end{align}
is the zero mode contribution that arises from target space momentum (the $n$ variable) and target space winding (the $m$ variable).  Here we have defined
\be\label{taudef}
\tau=\frac{\theta}{2\pi}+i\frac{\beta}{L},
\ee
used the usual Casimir renormalization
\be
\sum_{n=1}^\infty n =-\frac{1}{12},
\ee
introduced the Dedekind $\eta$ function
\be
\eta(\tau)\equiv e^{\frac{i\pi \tau}{12}}\prod_{n=1}^\infty\left(1-e^{2\pi i \tau n}\right),
\ee
and the second expression for $Z_0$ is a Poisson resummation of the first.
The partition function $Z_1$ is invariant under the modular transformation $\tau'=-\frac{1}{\tau}$, which follows from the $\eta$ function transformation  
\be\label{etatrans}
\eta\left(-\frac{1}{\tau}\right)=\sqrt{\frac{\tau}{i}}\eta(\tau)
\ee
and the fact 
\be
Z_0\left(-\frac{1}{\tau}\right)=\frac{Z_0(\tau)}{|\tau|},
\ee
which follows straightforwardly from our second expression for $Z_0$.

The various twisted partition functions \eqref{bossect} arising from background gauge fields for $\phi'=-\phi$ symmetry  are given explicitly by
\begin{align}\nonumber
Z_2&=\prod_{n=1}^\infty\frac{e^{-\frac{2\pi\beta n}{L}}}{\left(1+e^{2\pi i \tau n}\right)\left(1+e^{-2\pi i \ol{\tau} n}\right)}\\\nonumber
&=2\frac{|\eta(\tau)|}{|\Theta_2(\tau)|}\\\nonumber
Z_3&=2\prod_{n=1}^\infty\frac{e^{-\frac{2\pi\beta (n-1/2)}{L}}}{\left(1-e^{2\pi i \tau (n-1/2)}\right)\left(1-e^{-2\pi i \ol{\tau} (n-1/2)}\right)}\\\nonumber
&=2\frac{|\eta(\tau)|}{|\Theta_4(\tau)|}\\\nonumber
Z_4&=2\prod_{n=1}^\infty\frac{e^{-\frac{2\pi\beta (n-1/2)}{L}}}{\left(1+e^{2\pi i \tau (n-1/2)}\right)\left(1+e^{-2\pi i \ol{\tau} (n-1/2)}\right)}\\
&=2\frac{|\eta(\tau)|}{|\Theta_3(\tau)|}.
\end{align}
Here we have used the Casimir renormalization
\be
\sum_{n=1}^\infty(n-1/2)=\frac{1}{24},
\ee
and also that states of nonzero target space momentum or winding do not contribute to $Z_2$ and do not exist in the twisted sector.  On the other hand in the twisted sector there are two ground states corresponding to the points $\phi=0$ and $\phi=\pi R_\phi$, which give rise to the factors of $2$ in $Z_3$ and $Z_4$.
The Jacobi $\Theta$ functions are given by
\begin{align}\nonumber
\Theta_2(\tau)&=\sum_{n=-\infty}^\infty e^{i\pi\tau(n+1/2)^2}=2e^{i\pi \tau/4}\prod_{n=1}^\infty\left(1-e^{2\pi i \tau n}\right)\left(1+e^{2\pi i \tau n}\right)^2\\\nonumber
\Theta_3(\tau)&=\sum_{n=-\infty}^{\infty}e^{i\pi\tau n^2}=\prod_{n=1}^\infty\left(1-e^{2\pi i \tau n}\right)\left(1+e^{2\pi i \tau(n-1/2)}\right)^2\\
\Theta_4(\tau)&=\sum_{n=-\infty}^{\infty}(-1)^n e^{i\pi\tau n^2}=\prod_{n=1}^\infty\left(1-e^{2\pi i \tau n}\right)\left(1-e^{2\pi i \tau(n-1/2)}\right)^2, \label{thetadef}
\end{align}
and have the modular transformations
\begin{align}\nonumber
\Theta_2(-1/\tau)&=\sqrt{\frac{\tau}{i}}\Theta_4(\tau)\\\nonumber
\Theta_3(-1/\tau)&=\sqrt{\frac{\tau}{i}}\Theta_3(\tau)\\
\Theta_4(-1/\tau)&=\sqrt{\frac{\tau}{i}}\Theta_2(\tau).\label{thetatrans}
\end{align}
The full partition function once we gauge the symmetry by summing over background gauge fields is
\be
Z_{gauged}=\frac{Z_0}{2|\eta(\tau)|^2}+\frac{|\eta(\tau)|}{|\Theta_2(\tau)|}+\frac{|\eta(\tau)|}{|\Theta_4(\tau)|}+\frac{|\eta(\tau)|}{|\Theta_3(\tau)|},
\ee
which is indeed modular-invariant.

For the free fermion the untwisted partition function is
\begin{align}\nonumber
Z_{1}&=2\left|\prod_{n=1}^\infty e^{\frac{\pi \beta}{L}n}\left(1+e^{2\pi i n \tau}\right)\right|^2\\
&=\frac{|\Theta_2(\tau)|}{|\eta(\tau)|},
\end{align}
where the factor of $2$ arises from a zero mode consisting of a single qubit on which $\frac{1}{L}\int_0^L dx\psi_1(x)$ and $\frac{1}{L}\int_0^L dx\psi_2(x)$ act as $\sigma_x/\sqrt{2L}$ and $\sigma_y/\sqrt{2L}$ respectively.  We gauge fermion parity by summing over the various twisted partition functions \eqref{fermsect}.  $Z_2$ is actually zero since the insertion of fermion parity into the partition function causes a cancellation between the sectors built on top of the ground states with $\sigma_z=\pm 1$. $Z_3$ and $Z_4$ are given by
\begin{align}\nonumber
Z_3&=\left|\prod_{n=1}^\infty e^{\frac{\pi \beta}{L}(n-1/2)}\left(1+e^{2\pi i (n-1/2)\tau}\right)\right|^2\\\nonumber
&=\frac{|\Theta_3(\tau)|}{|\eta(\tau)|}\\\nonumber
Z_4&=\left|\prod_{n=1}^\infty e^{\frac{\pi \beta}{L}(n-1/2)}\left(1-e^{2\pi i (n-1/2)\tau}\right)\right|^2\\
&=\frac{|\Theta_4(\tau)|}{|\eta(\tau)|}.
\end{align}
Thus we have
\be
Z_{gauged}=\frac{1}{2}\left(\frac{|\Theta_2(\tau)|}{|\eta(\tau)|}+\frac{|\Theta_3(\tau)|}{|\eta(\tau)|}+\frac{|\Theta_4(\tau)|}{|\eta(\tau)|}\right),
\ee
which is indeed modular-invariant via \eqref{etatrans} and \eqref{thetatrans}.

\section{Regulated twisted background for $\T$}\label{regalg}
In this example we construct a discretized version of a scalar field on the Lorentzian mobius strip, consisting of three oscillators $y_1(t)$, $y_2(t)$, and $y_3(t)$ on a circle obeying the boundary condition $y_1(t)=y_3(-t)$.  Using this boundary condition to eliminate $y_3$, the equations of motion are
\begin{align}\nonumber
\ddot{y}_1(t)&=y_2(t)+y_2(-t)-2y_1(t)\\
\ddot{y}_2(t)&=y_1(t)+y_1(-t)-2y_2(t).
\end{align}
The general solution of these equations is
\be
\begin{pmatrix} y_1(t) \\ y_2(t) \end{pmatrix}=A\begin{pmatrix} 1 \\1\end{pmatrix}+B\begin{pmatrix} 1 \\ -1\end{pmatrix}\cos(2t)+C\begin{pmatrix}1\\-1\end{pmatrix}\sin(\sqrt{2}t)+D\begin{pmatrix} 1 \\1 \end{pmatrix}\sin(\sqrt{2}t), 
\ee
and as before we can write the parameters in terms of the initial data as
\begin{align}\nonumber
A&=\frac{1}{2}\left(y_1(0)+y_2(0)\right)\\\nonumber
B&=\frac{1}{2}\left(y_1(0)-y_2(0)\right)\\\nonumber
C&=\frac{1}{2\sqrt{2}}\left(\dot{y}_1(0)-\dot{y}_2(0)\right)\\
D&=\frac{1}{2\sqrt{2}}\left(\dot{y}_1(0)+\dot{y}_2(0)\right).
\end{align}
Imposing canonical commutation relations for $y_1$ and $y_2$, the non-vanishing commutators are
\begin{align}\nonumber
[A,D]&=\frac{i}{2\sqrt{2}}\\
[B,C]&=\frac{i}{2\sqrt{2}}.
\end{align}
We can also express this algebra classically in terms of the symplectic form:
\begin{align}\nonumber
\Omega&=\delta \dot{y}_1\wedge \delta y_1+\delta\dot{y}_2\wedge \delta y_2\\
&=2\sqrt{2}\left(\delta C\wedge \delta B+\delta D\wedge \delta A\right).
\end{align}

We can also consider a four-oscillator example with $y_1(t)=y_4(-t)$, which has equations of motion
\begin{align}\nonumber
\ddot{y}_1(t)&=y_2(t)+y_1(-t)-2y_1(t)\\\nonumber
\ddot{y}_2(t)&=y_1(t)+y_3(t)-2y_2(t)\\
\ddot{y}_3(t)&=y_2(t)+y_1(-t)-2y_3(t).
\end{align}
The general solution of the equations of motion is
\begin{align}\nonumber
\begin{pmatrix} y_1(t)\\y_2(t)\\y_3(t)\end{pmatrix}=&A\begin{pmatrix} 1\\1\\1\end{pmatrix}+B\begin{pmatrix} -1\\1\\1\end{pmatrix}\cos(\sqrt{2}t)+C\begin{pmatrix} 1\\-2\\1\end{pmatrix}\cos(\sqrt{3}t)\\
&+D\begin{pmatrix} 1\\2\\1\end{pmatrix}\sin(t)+E\begin{pmatrix} 1\\1\\-1\end{pmatrix}\sin(\sqrt{2}t)+F\begin{pmatrix} 1\\-1\\1\end{pmatrix}\sin(2t),
\end{align}
which we can invert to find
\begin{align}\nonumber
A&=\frac{1}{6}\left(3y_1(0)+2y_2(0)+y_3(0)\right) \qquad B=\frac{y_3(0)-y_1(0)}{2} \qquad C=\frac{y_3(0)-y_2(0)}{3}\\
D&=\frac{\dot{y}_2(0)+\dot{y}_3(0)}{3}\qquad E=\frac{\dot{y}_1(0)-\dot{y}_3(0)}{2\sqrt{2}}\qquad F=\frac{3\dot{y}_1(0)-2\dot{y}_2(0)+\dot{y}_3(0)}{12}.
\end{align}

\section{Winding of frames and the Euler characteristic}\label{windingapp}
In this appendix we derive the formula \eqref{winding}, which relates the the Euler characteristic of any closed oriented two-dimensional surface $\Sigma$ to the winding of conformal frames in patches.  This formula is closely related to the Poincare-Hopf theorem (see e.g. \cite{bott1982differential}), but our treatment will be self-contained.  We begin with the Gauss-Bonnet formula, which says that
\be
\chi(\Sigma)=\int_\Sigma T,
\ee
where 
\be
T=\frac{R}{4\pi}\epsilon
\ee
is the Euler density with $\epsilon$ being the volume form and $R$ being the Ricci scalar.  We can introduce a triangulation of $\Sigma$ whose faces we'll call $V_\alpha$.  The intersections of different faces are the precisely the edges of the triangulation, which we will denote as $V_{\alpha\beta}$ and assign an orientation such that
\be
\partial V_{\alpha}=\sum_\beta V_{\alpha\beta}.
\ee
Similarly the triple intersections of these faces are vertices $V_{\alpha\beta\gamma}$, which we orient so that
\be
\partial V_{\alpha\beta}=\sum_\gamma V_{\alpha\beta\gamma}.
\ee
In each patch $V_\alpha$ we can go to conformal gauge, putting the metric in the form
\be
ds^2=e^{2\phi_\alpha}dz_\alpha d\ol{z}_\alpha,
\ee
in terms of which we have
\be
T=dT_\alpha
\ee
with 
\be
T_\alpha=-\frac{1}{2\pi}\star d\phi_\alpha.
\ee
The different patches are related by holomorphic transformations
\be
z_\alpha=f_{\alpha\beta}(z_\beta),
\ee
so the conformal factors in different patches are related as
\be
\phi_\alpha\left(f_{\alpha\beta}(z_\beta),\ol{f}_{\alpha\beta}(\ol{z}_\beta)\right)-\phi_\beta(z_\beta,\ol{z}_\beta)=\frac{1}{2}\log |f_{\alpha\beta}'(z_\beta)|^2.
\ee
Noting that $\star dz_\beta=idz_\beta$ and $\star d \ol{z}_\beta=-id\ol{z}_\beta$, we have
\be
T_\alpha-T_\beta=dT_{\alpha\beta}
\ee
with
\be\label{EulerTab}
T_{\alpha\beta}(z_\beta,\ol{z}_\beta)=-\frac{i}{4\pi}\log\left(\frac{f_{\alpha\beta}'(z_\beta)}{\ol{f}_{\alpha\beta}'(\ol{z}_\beta)}\right).
\ee
In a triple overlap we must have
\be
f_{\alpha\beta}(f_{\beta\gamma}(f_{\gamma\alpha}(z_\alpha)))=z_\alpha,
\ee
so taking the derivative we find
\be
f'_{\alpha\beta}f'_{\beta\gamma}f'_{\gamma\alpha}=1.
\ee
Defining
\be
f'_{\alpha\beta}=r_{\alpha\beta}e^{i\theta_{\alpha\beta}},
\ee
we see that we must have
\be
r_{\alpha\beta}r_{\beta\gamma}r_{\gamma\alpha}=1
\ee
and
\be
\theta_{\alpha\beta}+\theta_{\beta\gamma}+\theta_{\gamma\alpha}=2\pi w_{\alpha\beta\gamma}
\ee
for some integers $w_{\alpha\beta\gamma}$.  Substituting into \eqref{EulerTab} we then have
\be
T_{\alpha\beta}=\frac{1}{2\pi}\theta_{\alpha\beta}.
\ee
This then tells us that we have
\be
T_{\alpha\beta}-T_{\alpha\gamma}+T_{\beta\gamma}=w_{\alpha\beta\gamma}.
\ee
We can then use these formulas to evaluate the Euler characteristic:
\begin{align}\nonumber
\chi(\Sigma)&=\int_\Sigma T\\\nonumber
&=\sum_\alpha\int_{V_\alpha}dT_\alpha\\\nonumber
&=\sum_\alpha\int_{\partial V_\alpha} T_\alpha\\\nonumber
&=\sum_{\alpha<\beta}\int_{V_{\alpha\beta}}(T_\alpha-T_\beta)\\\nonumber
&=\sum_{\alpha<\beta}\int_{\partial V_{\alpha\beta}}T_{\alpha\beta}\\\nonumber
&=\sum_{\alpha<\beta<\gamma}\int_{V_{\alpha\beta\gamma}}\left(T_{\alpha\beta}-T_{\alpha\gamma}+T_{\beta\gamma}\right)\\
&=\sum_{\alpha<\beta<\gamma}\int_{V_{\alpha\beta\gamma}}w_{\alpha\beta\gamma}.
\end{align}
This method gives a rather general way of understanding topological terms in field theories, which will be developed further elsewhere.  
\bibliographystyle{jhep}
\bibliography{bibliography}
\end{document}